\documentclass{aastex}
\usepackage{spr-astr-addons}
\usepackage{url}
\usepackage{psfig}   
\usepackage{subfigure}
\usepackage{graphicx,color,dcolumn}

\RequirePackage{color}

\begin{document}

\title{Magnetic field decay of magnetars in supernova remnants}
\slugcomment{Not to appear in Nonlearned J., 45.}
\shorttitle {magnetic fields' decay  }
\shortauthors{Z. F. Gao et al.}
\author{Z. F. Gao \altaffilmark{1,2,3}}
\altaffiltext{1}{Xinjiang Astronomical Observatory, CAS, 150, Science 1-Street, Urumqi Xinjiang, 830011,China zhifu$_{-}$gao@uao.ac.cn}
\altaffiltext{2}{Key Laboratory of Radio Astronomy, Chinese Academy of Sciences£¬Nanjing, 210008, China}
\altaffiltext{3}{Graduate University of the Chinese Academy of Sciences, 19A Yuquan Road, Beijing, 100049, China}
\author{ Q. H. Peng \altaffilmark{4}}
\altaffiltext{4}{Department of Astronomy, Nanjing University, Nanjing, 210093. China}
\author{ N. Wang \altaffilmark{1,2}}
\altaffiltext{1}{Xinjiang Astronomical Observatory, CAS, 150, Science 1-Street, Urumqi Xinjiang, 830011, China}
\altaffiltext{2}{Key Laboratory of Radio Astronomy, Chinese Academy of Sciences£¬Nanjing, 210008, China}
 \author{ J. P. Yuan \altaffilmark{1,2}}
\altaffiltext{1}{Xinjiang Astronomical Observatory, CAS, 150, Science 1-Street, Urumqi Xinjiang, 830011, China}
\altaffiltext{2}{Key Laboratory of Radio Astronomy, Chinese Academy of Sciences£¬Nanjing, 210008, China}
\begin{abstract}
In this paper, we modify our previous research carefully, and derive a new expression
of electron energy density in superhigh magnetic fields.  Based on our improved
model, we re-compute the electron capture rates and the magnetic fields'
evolutionary timescales $t$ of magnetars.  According to the calculated results,
the superhigh magnetic fields may evolve on timescales $\sim (10^{6}-10^{7})$ yrs
for common magnetars, and the maximum timescale of the field decay, $t\approx 2.9507
\times 10^{6}$ yrs, corresponding to an initial internal magnetic field $B_{\rm 0}=
3.0 \times 10^ {15}$ G and an initial inner temperature  $T_{\rm 0}=
2.6 \times 10^ {8}$ K. Motivated by the results of the neutron star-supernova
remnant(SNR) association of Zhang \& Xie(2011), we calculate the maximum $B_{\rm 0}$
of magnetar progenitors, $B_{\rm max}\sim (2.0\times 10^{14}-2.93 \times 10^{15})$ G when $T_{\rm 0}=
2.6 \times 10^ {8}$ K. When $T_{\rm 0}\sim 2.75 \times 10^ {8}-~1.75 \times 10^ {8}$ K, the
maximum $B_{\rm 0}$ will also be in the range of $\sim 10^{14}-10^{15}$ G,
not exceeding the upper limit of magnetic field of a
magnetar under our magnetar model. We also investigate the
relationship between the spin-down ages of magnetars and the ages of their SNRs, and
explain why all AXPs associated with SNRs look older than their real ages, whereas
all SGRs associated with SNRs appear younger than they are.
\end{abstract}

\keywords{Magnetar.\ Electron capture rate.\ Supernova remnant.\ and Superhigh magnetic field}

\section{Introduction}
Recent developments have shown that a substantial fraction of newly
born stars have magnetic field strengths in excess of the quantum
critical value, $B_{\rm cr}=4.414 \times 10^{13}$ G, above which the
effect of Landau quantization on the transverse electron motion becomes
considerable\citep{Ternov65}.  We divide them into two glasses$-$ Soft
Gamma-ray Repeaters(SGRs) and Anomalous X-ray Pulsars (AXPs) through
the studies of their emission mechanisms. The SGRs and AXPs are currently
considered to be `magnetars', powered by extremely strong magnetic fields,
rather than by their spin-down energy loss, as is the case for common
radio pulsars \citep{Duncan92, Thompson93, Thompson96}.

A supernova remnant(SNR) is an expanding diffuse gaseous nebula that
results from the explosion of a massive star. To date, there are 23
detected magnetar candidates: 11 SGRs (7 confirmed), and 12 AXPs (9
confirmed).  Of the magnetar candidates, 4 SGRs and 5 AXPs (more than
a third) are associated with the known SNRs, suggestive of an origin
in massive star explosions \citep{Gaensler01, Marsden01, Allen04,
Mereghetti08}.  Estimations indicate that about $10\%$ of supernova
explosions may lead to a magnetar \citep{Kouveliotou94}. If magnetars
are from core-collapse supernovae, as estimated above, then their
magnetic fields could have been inherited from their progenitors
\citep{Horiuchi08}.

The SGRs, characterized primarily by their occasional repeating bursts
of soft $\gamma$-rays, have spin periods $(5\sim 8)$ s, positive period
derivatives, and persistent soft X-ray luminosities $\sim 10^{35}$
erg~s$^{-1}$ \citep{Duncan00, Marsden01}. There are four known
SGRs associated with SNRs: O526-66, 1900+14, 1806-20 and 1627-41. SGR 0526-66
has been associated with N49 in the Large Magellanic Cloud \citep{Kulkarni03}.
SGR 1806-20 and SGR 1627-41 apparently lie in G10.0-0.3 \citep{Kulkarni93}
and  G33.70-01 \citep{Corbel99}, respectively. SGR 1900+14 is associated with
G42.8+0.6 \citep{Hurley99}.  However, another SNR G43.9+1.6 also falls within
the error box of SGR 1900+14 \citep{Vasisht94}, so a great deal of effort will
be required to investigate whether  this magnetar is stably associated with
G42.8+0.6.

The AXPs, so called due to their high X-ray luminosities, $10^{34}\sim 10^{36}$
erg~s$^{-1}$, the lack of evidence of binary companions \citep{van95, Duncan96}.
To date, there are five AXPs are located near the centers of SNRs: 1E 2259+586
in CTB 109 \citep{Fahlman81}, 1E 1841-045 in Kes 73 \citep{Sanbonmatsu92}, 1E
1547.0-5408 in G327.24-0.13 \citep{Camilo07}; CXOU J171405.7-381031 in CTB37B
\citep{Aharonian08} and AX J1845-0258 in G29.6+0.1 \citep{Gaensler99}.

Due to a very little number of magnetars (only 16 confirmed currently), some
puzzles concerning supernova progenitors and their birth events that confront
us seem inevitable.  However, a few confirmable magnetar/SNR associations still
remain undisputed \citep{Gaensler01, Marsden01, Allen04}. These associations are
confirming that a magnetar candidate was formed in a supernova explosion, and is
thus thought to be the collapsed core of a massive star, which is likely to be
a neutron star(NS).  Since a NS and its associated SNR are from the same
explosion, they should have the same age \citep{Zhang11}.  All the known SNRs of
both AXPs and SGRs are comparatively young, $t_{\rm SNR}\sim$ several kyr (see
Sec.3), which infers that magnetars are also very young objects \citep{Shull89,
Vasisht97}.

Over the last decades, radio pulsars(here, observed as normal radio pulsars) with
the possible magnetic field evolutionary timescale, $t\sim10^{6}-10^{7}$ yrs, have
been studied extensively using the statistical distribution in the $P-\dot{P}$ diagram
\citep{Ostriker69, Gunn70, Shull89, Narayan90, Han97, Ibrahim04, Aguilera08, Ridley10,
Zhang11}. Maybe radio pulsars were born with different initial external circumstances
and initial internal conditions, these pulsars have experienced different evolutionary
routes: the magnetic fields remain nearly constant in about half of them, decrease
rapidly in others, or increase in very few pulsars \citep{Han97, Zhang11}.

In order to explain the evolution of superhigh magnetic fields inside magnetars,
different models have been proposed recently, as partly listed below.  According
to the twisted magnetospheres model, the unwinding of the internal field shears
the  star's crust, the rotational crustal motions generally provide a source
of helicity for the external magnetosphere by twisting the magnetic fields which are
anchored to the star's surface, drive currents outside a magnetar, and generate X-ray
emissions\citep{Thompson00}.  The sudden crustal fractures (or starquakes) caused
by unbearable stress can provide plausible mechanisms for magnetar outbursts and giant
flares \citep{Thompson00, Thompson02}.  In addition, the overall evolution of SGR 1806
-20 in the years preceding the giant flare of December 27, 2004 seems to support some
predictions of this model \citep{Mereghetti05}.

In the thermal evolution model, the field could decay directly as a result of the
non-zero resistivity of the matter through Ohmic decay or ambipolar diffusion, or
indirectly as a result of Hall drift producing a cascade of the field to high wave
number components, which decay quickly through Ohmic decay \citep{Goldreich92,
Rheinhardt03, Pons06}; magnetic field decay can be a main source of internal heating
\citep{Pons09, Arras04}; the enhanced thermal conductivity in the strongly magnetized
envelope contributes to raise the surface temperature \citep{Heyl97, Heyl98}.  Based
on this model, the surface thermal temperature of a magnetar is estimated to be
$\sim (10^{5}-10^{6})$ K, which is basically consistent with the observations \citep{Heyl98,
Pons09}.

Although there are apparent advantages in some magnetar models, including the above
two magnetar models, substantial improvements must be made for these models. The main
disadvantages of these models can be summarized as follows: (1) not considering the
effects of anisotropic ${}^3P_{\rm 2}$ neutron superfluid (mainly in the outer core)
on the decay of superhigh magnetic fields; (2) not combining the ages of SNRs with the
timescales of magnetic fields' evolution; (3) universally adopting the most popular
assumption on the origin of superhigh magnetic fields of magnetars $-$`$\alpha-\Omega$
dynamo' \citep{Duncan92, Thompson93}, which is a mere assumption laking observational
support \citep{Gao11a, Gao11b, Gao11c, Gao11d}(hereinafter Paper 1, Paper 2, Paper 3
and Paper 4, respectively).

Unlike other magnetar models, we propose that superhigh magnetic fields of magnetars
originate from the induced magnetic fields below a critical temperature, and the
maximum field strength is $\sim (3.0-4.0)\times 10^{15}$ G \citep{Peng07, Peng09}.  In
the initial stage of our magnetar model, the major conclusions are briefly summarized
as following: In Paper 1, we numerically simulated the whole process of electron capture
(EC); in order to calculate the effective electron capture rates, $\Gamma_{\rm eff}$,
we introduced the Landau level effect coefficient, $q$, whose magnitude is evaluated
to be $\sim 10^{-18}$ by comparing the observed magnetars' soft X-ray luminosities $L_
{\rm X}$ with the calculated values of $L_{\rm X}$. In Paper 2, superhigh magnetic fields
give rise to an increase in the electron
Fermi energy $E_{\rm F}({\rm e})$, which will induce EC inside a magnetar. The ${}^3P
_{\rm 2}$ Cooper pairs with the maximum binding energy of 0.048 MeV \citep{Elgar96}
will be destroyed by the outgoing high-energy EC neutrons. Then the magnetic moments
of the ${}^3P_{\rm 2}$ Cooper pairs destroyed are no longer arranged in the paramagnetic
direction, so the superhigh magnetic fields produced by the aligned magnetic moments of
the ${}^3P_{\rm 2}$ Cooper pairs will disappear gradually.  Combining parameter $q$
with ${}^3P_2$ anisotropic neutron superfluid theory yields a second-order differential
equation for superhigh magnetic fields $B$ and  their evolution timescales $t$.  In
Paper 3, by introducing the Dirac $\delta$-function, we deduced a general formula for
$E_{\rm F}({\rm e})$, which is suitable for extremely intense magnetic fields. In Paper
4, we presented the mechanism for the magnetar soft X/$\gamma$-ray emission, numerically
simulated the process of magnetar cooling and magnetic field decay, and then computed
$L_{\rm X}$ of magnetars by introducing two important parameters: Landau level-superfluid
modified factor $\Lambda$ and effective X/$\gamma$-ray coefficient $\zeta$.

In this work, we re-examine our previous research carefully, and find that the expression
of energy state density $\rho_{\rm e}$ of electrons in superhigh magnetic fields, as
that of $E_{\rm F}({\rm e})$, should be derived in circular cylindrical coordinates
rather than in spherical coordinates, because the Landau column becomes a very long
and narrow cylinder along the magnetic field.  In Appendix B of this article, we modify
the expression of $E_{\rm F}({\rm e})$, derive a new formula of electron energy state
density in superhigh magnetic fields, improve the calculated results of $L_{\rm X}$
and $\zeta$, and compare these results with those of Paper 4. Apart from the
modifications above, the expression of $\Gamma$, together with the second-order
differential equation for $B$ and $t$ in Paper 2, must be improved. The main reasons
are as follows:\\
\begin{enumerate}
\item In the interior of a magnetar, the process of EC is a relatively independent
process, and is irrelevant to the luminosity $L_{\rm X}$, the values of $\Gamma$ are
completely determined by inner physical properties, e.g., magnetic field strength,
matter density, temperature, neutron superfluid and so on.

\item  Since the magnitude of $q$ can be estimated by the observed luminosities $L_
{\rm X}$ (see Paper 1), in fact, $q$ have included the influences of all the
following factors: neutron superfluid's restraining effect, the fractions of all
particles participating in EC reaction, thermal energy loss, energy conversion
efficiency, and gravitation redshift, accidentally.  However, if the influences of
the above factors are considered, the real value of $q$ will be far less than that
of Paper 1, therefore, $q$ is no longer used by our improved model. In Paper 4, $q$
has been replaced by two important parameters: Landau level-superfluid modified
factor, $\Lambda$, and effective X/$\gamma$-ray coefficient, $\zeta$ when calculating
$L_{\rm X}$.

\item  In our previous work (Papers 1-4), all the expressions of $\Gamma$, however, do
not make use of the quantity $\Lambda$ and the new expression of $\rho_{\rm e}$ in
superhigh magnetic fields (see Appendix B), so these expressions of $\Gamma$ are not
consistent with the actual circumstances inside magnetars. All of these strongly suggest
a necessity of reconsidering $\Gamma$ and the equation of $B$ and $t$.
\end{enumerate}

In 2011, Shuang-Nan Zhang and Yi Xie published an article titled `Magnetic field decay
makes NSs look older than they are', which showed convincing evidence of magnetic field
decay in some young NSs, and reasoned that the magnetic field decay can change
substantially their spinning behaviors, and as a result these NSs appear much older than
their real ages \citep{Zhang11}(hereinafter referred to as ZX2011).  According to ZX2011,
a NS and its associated SNR should have the same age, i.e., $t_{\rm Real} = t_{\rm SNR}$;
the NS's spin-down or characteristic age can be generally expressed as, $t_{\rm Spin}=
P/(n-1)\dot{P}$, where $P$, $\dot{P}$ and $n$ are its spin period, period derivative and
braking index, respectively; however, $n$ always deviates from the the value ($n = 3$)
expected for pure magnetic dipole radiation model. The authors supposed that, $n\gg 3$
is required for neutrons with $
t_{\rm Spin}\ll t_{\rm SNR}$ if there is no significant magnetic field decay (note: this
case is not believed to be plausible by authors), $n< 3$ makes the spin-down age of a NS
even longer than assuming $n= 3$, and in this case more or even all all NSs have
$t_{\rm Spin}\ll t_{\rm SNR}$; for any reasonable values of $n$, at least some of these
NSs must have experienced significant dipole magnetic field decay.  Furthermore, magnetic
field decay dominated by the ambipolar diffusion has been investigated, and the core and
surface temperatures of a NS have been estimated, whose results are agreed qualitatively
with observations \citep{Pons09}.  However, authors did not provide the observation data
of magnetars, and thus omitted to explain why all AXPs associated with SNRs look older
than their real ages, whereas all SGRs associated with SNRs appear younger than they are.

The remainder of this paper is organized as follows. In Sec.2, by introducing two different
types of electron energy state density, the electron capture rates $\Gamma$ in superhigh
magnetic fields are calculated, and the calculated results are compared with each other. In Sec.3.1, the
differential equation of $B$ and $t$ is modified, and the values of $t$ are re-computed.
In Sec.3.2, the maximum initial internal fields of magnetar progenitors are computed taking
$t_{\rm Real}= t_{\rm SNR} $ proposed by ZX2011 as the starting point. In Sec.3.3, the
relationship between the spin-down ages of magnetars and the ages of their SNRs are
investigated. In Sec.4, a brief summary is given. In Appendix A, an important assumption
on the ${}^3P_{\rm 2}$ neutron Cooper pairs is presented, and several corrections and
improvements in our models are presented in Appendix B.
\section{Electron capture rates in superhigh magnetic fields}
Since the quantized microstates don't exist in the momentum (or energy) space between
the $n$-th and $(n+1)$-th Landau level, the Dirac $\delta$-function must be taken into
account when calculating $E_{\rm F}({\rm e})$ in superhigh magnetic fields. From Appendix
B, a concise expression for $E_{\rm F}({\rm e})$ in superhigh magnetic fields is of the form,
\begin{equation}
E_{\rm F}({\rm e})\simeq 43.44(\frac{B}{B_{\rm cr}})^{\frac{1}{4}}(\frac{\rho}{\rho_{0}}\frac{Y_{\rm e}}{0.0535})
^{\frac{1}{4}}~~~~~~~~~\rm MeV~~.
\end{equation}
where $\rho_{\rm 0}= 2.8\times 10^{14}$ g cm$^{3}$ is the standard nuclear density.
In order to calculate the EC rate, $\Gamma$ in a magnetar, we concentrate on
non-relativistic, degenerate nuclear matter and super-relativistic, degenerate
electrons.  In the case of $0.5 \rho_{\rm 0}\leq \rho \leq  2 \rho_{\rm 0}$, the
following expressions hold approximately: $E_{\rm F}^{'}({\rm n})=60(\rho/\rho_
{\rm 0})^{\frac{2}{3}}$ MeV and $E_{\rm F}^{'}({\rm p})= 1.9(\rho/\rho_{\rm 0})
^{\frac{4}{3}}$ MeV, where $E_{\rm F}^{'}({\rm n})$ and $E_{\rm F}^{'}({\rm p})$
are the neutron Fermi kinetic energy and the proton Fermi kinetic energy,
respectively \citep{Shapiro83}.  In this paper, for convenience, we set $\rho= \rho_
{\rm 0}$ and the electron fraction $Y_{\rm e}=0.0535$ in all the following
calculations.  This choice yields the threshold energy of EC reaction, $Q =E_{\rm F}
({\rm n})-E_{\rm F}({\rm p})$= 59.39 MeV. The range of $B$ is assumed to be $B\sim
(B_{\rm th} \sim 3.0 \times 10^{15}$ G), where $B_{\rm th}=1.5423 \times 10^{14}$ G
is the threshold magnetic field of EC reaction, corresponding to $E_{F}({\rm e})$=
59.39 MeV. Thus, the range of $E_{\rm e}$ is ($Q \sim E_{\rm F}({\rm e})$). By
employing energy conservation via $E_{\nu}+ E_{\rm n}=E_{\rm e}+E_{\rm p}$, the
Fermi energy of neutrinos, $E_{\rm F}(\nu)$, can be calculated by
\begin{equation}
E_{\rm F}(\nu) = E_{\rm F}({\rm e})- Q = E_{\rm F}({\rm e})- 59.39 ~~~~~\rm MeV~.
\end{equation}

According to our point of view, once the energies of electrons near the Fermi surface
exceed $Q$, the EC reaction will dominate (see Paper 2 and Paper 4).  From Appendix B,
the energy state density of electrons in superhigh magnetic fields is of the form:
\begin{eqnarray}
&&\rho_{\rm e}\simeq \frac{4}{3}\frac{\pi}{B^{*}}(\frac{m_{\rm e}c}{h})^{3}
\frac{1}{m_{\rm e}c^{2}}[(\frac{E_{F}({\rm e})}{m_{\rm e}c^{2}})^{2}-1-(\frac{E_{\rm e}}{m_{\rm e}c^{2}})^{2}]^{\frac{3}{2}}\nonumber\\
&&=\frac{1}{3B^{*}(2\pi^{2}\hbar^{3}c^{3})}\frac{1}{m_{\rm e}c^{2}}[E_{F}^{2}({\rm e})- 0.261 - E_{\rm e}^{2}]^{\frac{3}{2}}~~.
\end{eqnarray}
where $B^{*}$ is a non-dimensional magnetic field, defined as $B^{*}= B/B_{\rm cr}$.
Since neutrinos/antineutrinos are uncharged, the energy state density of neutrinos
(antineutrinos) remains unchanged,
\begin{equation}
\rho_{\rm {\nu}} = \frac{(E_{\rm e}-Q)^{2}}{2\pi^{2}\hbar^{3}c^{3}}~~.
\end{equation}

The EC rate $\Gamma$, defined as the number of electrons captured by one proton per
second, can be computed using the standard charged-current $\beta$-decay theory
\citep{Shapiro83}. However, for each degenerate species, only a fraction ($\sim
kT/E_{\rm F}({\rm i})$) of particles near the Fermi surface can effectively contribute to
$\Gamma$.  In Paper 4, we introduced the `Landau level-superfluid modified factor' $\Lambda$,
\begin{equation}
\Lambda=\frac{(kT)^{4}Exp(-\Delta_{\rm max}({}^3P_{\rm 2})/kT)}{E_{\rm F}^{'}({\rm n})E_{\rm F}^{'}({\rm p})E_{\rm F}({\rm e})E_{\rm F}(\nu_{\rm e})}~.
\end{equation}
The initial conditions (e.g., $B, T$) of the magnetars
are likely to be very different, implying that their values of $\Lambda$ are
also different. For convenience, we assume a uniform initial magnetic field $B_{0}
=3.0\times 10^{15}$~G for magnetar progenitors. Since the process of EC is a
precess of magnetic field decay and inner cooling, when electrons are captured,
the numbers of particles participating in EC near the Fermi surfaces decrease, which
leads to a decrease in $\Lambda$. However, in the interior of a magnetar, the $\beta-$
decay and the inverse $\beta-$decay occur simultaneously as required by the charge
neutrality, so when B decays, the depleted protons and electrons are recruited many
times, which leads to only a small decrease in $Y_{\rm e}$ and $Y_{\rm p}$. In addition,
the electrons are super-relativistic and degenerate, when the internal temperature falls,
the electron transition between Landau levels is not permitted, because the electrons can
be approximately treated as a zero-temperature Fermi gas. Thus, the value of $\Lambda$
decreases very slowly. In order to obtain a fitting function of $\Lambda$, $B$ and $T$,
we numerically simulate the inner cooling and the magnetic field decay. Since the internal
temperature of a magnetar is $\sim 10^{8}$ K \citep{Yakovlev01}, and the maximum
initial temperature (not including the inner core temperature) cannot exceed the critical
temperature of the ${}^3P_{\rm 2}$ Cooper pairs $T_{\rm cn}\sim \Delta_{\rm max}({}^3P_{\rm 2})
/kT\sim 2.78\times10^{8}$ K (Paper 4), we can arbitrarily  assume $T_{0}$ to be $2.60\times
10^{8}$ K, corresponding to an initial value of $\Lambda\sim 3.198 \times10^{-14}$. Then, we gain
 \begin{eqnarray}
&&\Lambda(B_{0}, T_{0})= \Lambda(T_{0})=3.198\times 10^{-14}(\frac{T_{0}}{2.6 \times 10^{8}{\rm K}})^{4}\nonumber\\
&&Exp[\frac{-0.048{\rm MeV}}{k}(\frac{1}{T_{0}}-\frac{1}{2.6\times 10^{8}{\rm K}})]~.
\end{eqnarray}
When $B$ decreases from $B_{0}$ to $B_{\rm th}$, the ratio of $|\frac{\Delta T}{\Delta B}|_{\rm max}\sim
|\frac{(2.78\times 10^{8}-1.0\times 10^{8})\rm K}{(3.0\times 10^{15}-1.5423\times 10^{14})\rm G}|
\sim 5.33\times 10^{-8}$~K/G.  When simulating numerically, the assumed value of $|\frac{\Delta T}
{\Delta B}|$ cannot be too high or too small, otherwise the internal temperature $T$ drops wildly (eg.,
$T\ll 10^{7}$ K) or insignificantly.  According to the analysis above, we arbitrarily set $B_{0}=3.0
\times 10^{15}$, $T_{0}= 2.60\times 10^{8}$~K and $\Lambda(B_{0}, T_{0})=3.198\times 10^{-14}$ (these
specific values are representative of the initial conditions  encountered), and $T$ is decreased by
step $\Delta T= 3.5\times 10^{4}$ K. The results of numerical simulations
are shown in Fig.1.
\begin{figure*}[htb]
\begin{center}
\begin{tabular}{cc}
\scalebox{0.89}{\includegraphics{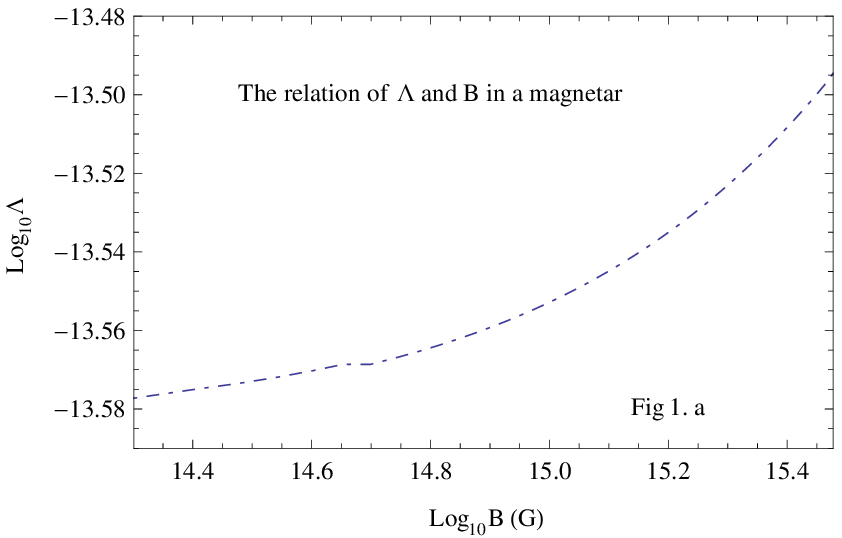}}&\scalebox{0.91}{\includegraphics{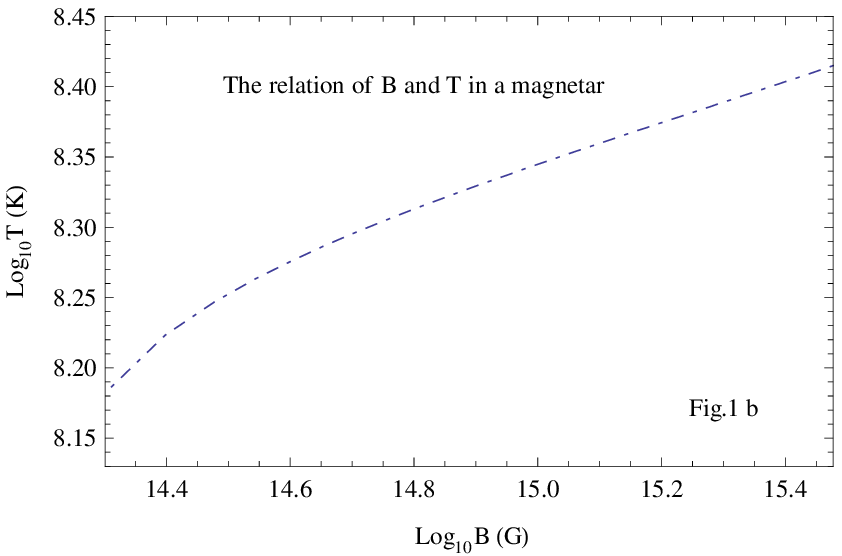}}\\
(a)&(b)\\
\end{tabular}
\end{center}
\caption{$\Lambda$ as a function of magnetic field $B$. The range of $B$ is about ($3.0\times 10^{15}- 2.0 \times
10^{14}$) G when $T_{0}= 2.60\times 10^{8}$ K.
 \label{fg:Fig.1}}
\end{figure*}
From Fig.1 a and Fig.1 b, both $\Lambda$ and $T$ decrease with decreasing $B$. When $B\sim 3.0\times 10^{15}- 2.0 \times
10^{14}$) G, $\Lambda\sim 3.198\times 10^{-14}- 2.649\times 10^{-14}$. Base on the simulations above we gain
\begin{eqnarray}
&&\Lambda (B, T)\approx (2.60388\times 10^{-14}+~1.98233\nonumber\\
&& \times 10^{-30}B)(\frac{T_{0}}{2.6\times10^{8}\rm K})^{4}Exp[\frac{-0.048{\rm MeV}}{k}\nonumber\\
&&(\frac{1}{T_{0}}-\frac{1}{2.6\times10^{8}{\rm K}})],
\end{eqnarray}
where $B_{0}=3.0\times 10^{-15}$~G is used. In the same way, we obtain a diagram of $\Lambda$ as a
function of internal temperature $T$, as shown in Fig.2.
\begin{figure}[th]
\centering
  \vspace{0.5cm}
  \includegraphics[width=8.1cm,angle=360]{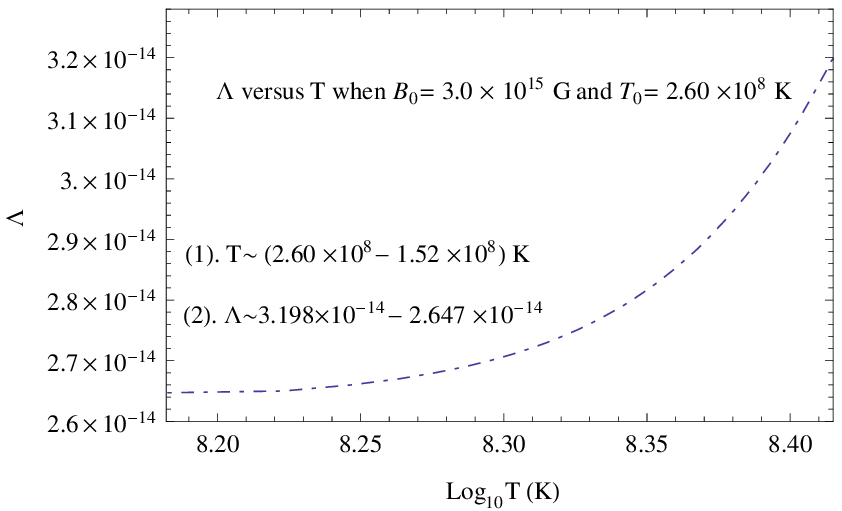}
  \caption{$\Lambda$ as a
function of internal temperature $T$. The range of $T$ is about ($2.60\times 10^{8}- 1.52\times
10^{8}$) K when $B_{0}= 3.0\times 10^{15}$ G.}
  \label{fg:Fig.2}
  \end{figure}
From Fig.2, when internal temperature $T$ drops from $2.60\times 10^{8}$ K to $1.52\times
10^{8}$ K, $\Lambda$ decreases, but its order of magnitude remains unchanged.

Using Eq.(7), we obtain the expression of $\Gamma$ in superhigh magnetic fields,
\begin{eqnarray}
&&\Gamma=\Lambda (B, T)\frac{2\pi}{\hbar}G_{\rm F}^{2}C_{\rm V}^{2}(1 + 3a^{2})\int_{Q}^{E_{F}({\rm e})}f_{\rm e}(1-f_{\nu})\nonumber\\
&&\times f_{\rm p}(1-f_{\rm n})\rho_{\nu} \rho_{\rm e}dE_{\rm e}\rho_{\nu}dE_{\nu}\delta(E_{\nu}+ Q - E_{\rm e})\nonumber\\
&&=\Lambda~\frac{2\pi G_{F}^{2}C_{V}^{2}(1 + 3a^{2})}{\hbar(2\pi^{2}\hbar^{3}c^{3})^{2}3m_{\rm e}c^{2}}
\frac{1}{B^{*}}\int_{Q}^{E_{F}({\rm e})}f_{\rm e}(1-f_{\nu})f_{\rm p}\nonumber\\
&&\times(1-f_{\rm n})[E_{F}^{2}({\rm e})-0.261-E_{\rm e}^{2}]^{\frac{3}{2}}(E_{\rm e}-Q)^{2})dE,
\end{eqnarray}
where $f({\rm j})=[Exp((E_{\rm j}-\mu_{\rm j})/kT)+ 1]^{-1}$ is the fraction of phase
space occupied at energy $E_{\rm j}$ (Fermi-Dirac distribution), factors of $(1-f_{\rm j})$
reduce the reaction rate, and are called `blocking factor', inside a NS, for neutrinos(antineutrinos),
$(1-f_{\nu})$= 1; for electrons, when $E_{\rm e} < E_{\rm F}({\rm e})$, $f_{\rm e}$=1, when $E_{\rm e}
 > E_{\rm F}({\rm e})$, $f_{\rm e}$ = 0; for neutrons, when $E_{k}({\rm n})< E_{\rm F}^{'}({\rm n})$,
$(1-f_{\rm n})$= 0, when $E_{k}({\rm n})> E_{\rm F}^{'}({\rm n})$, $(1-f_{\rm n})$= 1; for protons: when
$E_{\rm p} < E_{\rm F}({\rm p})$, $f_{\rm p}$= 1, when $E_{\rm p}> E_{\rm F}({\rm p})$,
$f_{\rm p}$= 0, so $f_{\rm e}(1-f_{\nu})f_{\rm p}(1-f_{\rm n})\simeq 1$ can be ignored in the latter
calculations; the other quantities have been defined in Paper 1, Paper 2 and Paper 4.  Eq.(6) is the
very expression of $\Gamma$ we have been looking for. Compared with Eqs.(3-4) of Paper 2, the advantages
of Eq.(6) mainly includes: (1) It reflects the effective or actual capture rates of electrons in superhigh
magnetic fields, which can be calculated directly using Eq.(6), rather than be modified by parameter $q$;
(2) It adopts the updated expressions of $\rho_{\rm e}$ and $E_{\rm F}({\rm e})$ in superhigh magnetic fields,
both of which are derived in circular cylindrical coordinates. In order to compare the results of $\Gamma$
in superhigh magnetic fields calculated by two different types of electron energy state density, we appeal
to the following expression,
\begin{eqnarray}
&&\Gamma^{'}=\Lambda(B, T)\frac{2\pi}{\hbar}G_{\rm F}^{2}C_{\rm V}^{2}(1+3a^{2})\int_{Q}^{E_{\rm F}({\rm e})}f_{\rm e}(1-f_{\nu})\nonumber\\
&& \times f_{\rm p}(1-f_{\rm n})\rho_{\rm e}dE_{\rm e}^{'}\rho_{\rm {\nu}}dE_{\rm {\nu}}\delta(E_{\rm {\nu}}+ Q- E_{\rm e})\nonumber\\
&&=\Lambda~\frac{2\pi}{\hbar}\frac{G_{\rm F}^{2}C_{\rm V}^{2}(1+3a^{2})}{(2\pi^{2}\hbar^{3}c^{3})^{2}}\int_{Q}^{E_{\rm F}(e)}f_{\rm e}(1-f_{\nu})\nonumber\\
&&\times f_{\rm p}(1-f_{\rm n})(E_{\rm e}^{2}- m^{2}_{\rm e}c^{4})^{\frac{1}{2}}E_{\rm e}(E_{\rm e}- Q)^{2}dE_{\rm e}~~,
\end{eqnarray}
where the most common and typical expression for electron energy state density $\rho_{\rm e}^{'}$
in a sphere symmetrical momentum space, $\rho_{\rm e}^{'}=\frac{4\pi p_{\rm e}^{2}}{h^{3}}\frac{dp_{\rm e}}
{dE_{\rm e}}= \frac{4\pi p_{\rm e}E_{\rm e}}{c^{2}h^{3}}$, is used.  Inserting $\frac{2\pi}{\hbar}
\frac{G_{\rm F}^{2}C_{\rm V}^{2}(1+3a^{2})}{(2\pi^{2}\hbar^{3}c^{3})^{2}}=0.018(\rm MeV)^{-5}~{\rm s}^{-1}$ and
$m_{\rm e}c^{2}$= 0.511 MeV into Eq.(8) and Eq.(9), we obtain the values of electron capture rate in superhigh
magnetic fields.  The calculation results are partly listed below in tabular form.
\begin{table*}[t]
\small
\caption{The calculated values of electron capture rates in superhigh magnetic fields  \label{tb1-1}}
\begin{tabular}{@{}crrrrrrr@{}}
\tableline
$ B$      & $E_{\rm F}({\rm e})$  &   $E_{\rm F}(\nu)$ &$ \langle E_{\rm n} \rangle $&$\langle E_{\rm \nu}\rangle$&$\Gamma$&$\Gamma^{'}$ & $K= \Gamma/\Gamma^{'}$\\
 (~G~)       &(~MeV~)      &(~MeV~)  &(~MeV~)      &(~MeV~)    &(~s$^{-1}$~) &(~s$^{-1}$~) &  \\
\tableline
1.8~$\times 10^{14}$ &61.73  &2.34  &61.06 &1.28 & 2.43~$\times 10^{-13}$ & 7.76~$\times 10^{-12}$ & 3.14~$\times 10^{-2}$\\
2.0~$\times 10^{14}$ &63.38  &3.99  &61.81 &2.18 & 2.50~$\times 10^{-12}$ & 4.01~$\times 10^{-11}$ & 6.24~$\times 10^{-2}$\\
2.5~$\times 10^{14}$&67.01   & 7.62 &63.44 &4.18 & 3.95~$\times 10^{-11}$ & 3.06~$\times 10^{-10}$ & 1.29~$\times 10^{-1}$ \\
3.0~$\times 10^{14}$ &70.14 &10.75  &64.84 &5.91 & 1.64~$\times 10^{-10}$ & 9.24~$\times 10^{-10}$  & 1.78~$\times 10^{-1}$\\
4.0~$\times 10^{14}$ &75.37  &15.98 &67.16 &8.82 & 8.07~$\times 10^{-10}$ & 3.43~$\times 10^{-9}$  & 2.35~$\times 10^{-1}$\\
5.0~$\times 10^{14}$ &79.69  &20.30 &69.07 &11.23 & 2.04~$\times 10^{-9}$ & 7.74~$\times 10^{-9}$   & 2.63~$\times 10^{-1}$\\
6.0~$\times 10^{14}$&83.41   &24.02 &70.71 &13.31 & 3.87~$\times 10^{-9}$ & 1.39~$\times 10^{-8}$  & 2.78~$\times 10^{-1}$\\
7.0~$\times 10^{14}$ &86.69  &27.30 &72.15 &15.15 & 6.24~$\times 10^{-9}$ & 2.19~$\times 10^{-8}$   & 2.84~$\times 10^{-1}$\\
8.0~$\times 10^{14}$&89.63   &30.24 &73.43 &16.81 & 9.08~$\times 10^{-9}$ & 3.17~$\times 10^{-8}$  & 2.86~$\times 10^{-1}$\\
9.0~$\times 10^{14}$&92.31   &32.92 &74.60 &18.32 & 1.24~$\times 10^{-8}$ & 4.33~$\times 10^{-8}$  & 2.85~$\times 10^{-1}$\\
1.0~$\times 10^{15}$&94.77   &35.38 &75.67 &19.71 & 1.60~$\times 10^{-8}$ & 5.61~$\times 10^{-8}$  & 2.83~$\times 10^{-1}$\\
1.5~$\times 10^{15}$&104.88  &45.49 &80.05 &25.44 & 3.90~$\times 10^{-8}$ & 1.48~$\times 10^{-7}$  & 2.64~$\times 10^{-1}$\\
2.0~$\times 10^{15}$&112.70 &53.31  &83.43 &29.88 & 6.78~$\times 10^{-8}$ & 2.78~$\times 10^{-7}$  & 2.43~$\times 10^{-1}$\\
2.5~$\times 10^{15}$&119.17 &59.78  &86.21 &33.57 & 1.01~$\times 10^{-7}$ & 4.47~$\times 10^{-7}$  & 2.26~$\times 10^{-1}$\\
2.8~$\times 10^{15}$& 122.59 &63.20 &87.67 &35.52 & 1.22~$\times 10^{-7}$ & 5.56~$\times 10^{-7}$  & 2.20~$\times 10^{-1}$\\
3.0~$\times 10^{15}$&124.72 &65.33  &88.59 &36.74 & 1.38~$\times 10^{-7}$ & 6.52~$\times 10^{-7}$  & 2.11~$\times 10^{-1}$\\
\tableline
  \end{tabular}
\end{table*}
The main results in Table 1 are summarized as follows: When the magnetic field $B\sim (3.0
\times 10^{15}- 1.8\times 10^{14}$) G, $\Gamma \sim(1.38 \times 10^{-7}- 2.43\times 10
^{-13})$~s$^{-1}$~and~$\Gamma^{'} \sim(6.52 \times 10^{-7}- 7.76 \times 10^{-12})$~s
$^{-1}$, respectively.  From Table 1, the values of $\Gamma$ are universally less than
those of $\Gamma^{'}$, and the ratio of $K=\Gamma/\Gamma^{'}$ is about the magnitude of
$10^{-1}-10^{-2}$. The possible explanations are as follows:\\
\begin{enumerate}

\item  Due to  the formation of Landau cylinder in the momentum space or the
quantization of Landau levels, the spherical symmetry in the momentum space
is broken by superhigh magnetic fields.  The formula of $\rho_{\rm e}$ and that of
$\rho_{\rm e}^{'}$ are derived in circular cylindrical coordinates and in
spherical coordinates, respectively.

\item  In the vicinity of the Fermi surface, the electrons with the same energy
$E$ could come from different Landau levels because the electrons are degenerate.
However, the electrons occupying lower Landau levels cannot be captured even if
their energies are higher than $Q$; for higher Landau levels, there still exist
some electrons with lower energies $E$($E< Q$) that are not captured. Thus, the
values of $\Gamma$ calculated by the expression of $\rho_{\rm e}$ in superhigh
magnetic fields are less than those of $\Gamma^{'}$ computed by the expression of
$\rho_{\rm e}^{'}$ in non-relativistic magnetic fields.

\item In Papers 1-2, in order to obtain the values of the effective electron capture rates $\Gamma
_{\rm eff}$ in superhigh magnetic fields, we introduce the Landau level effect
coefficient $q$. In our improved model, the following
factors: neutron superfluid's restraining effect, the numbers of all particles
participating in EC reaction, thermal energy loss, energy conversion efficiency and
so on, have been considered, so the quantity $q$~must be replaced by two important
parameters: $\Lambda$ and $\zeta$ (see Paper 4).  Actually, the quantity $q$
includes the effects of $\Lambda$ and $\zeta$. Thus, the  values of $K$
is far less than those of $q$ (see Papers 1-2).
\end{enumerate}
From Table 1, we obtain the diagram of $K$ as a function of $B$, shown as in Fig.3.
\begin{figure}[th]
\centering
  \vspace{0.5cm}
  \includegraphics[width=8.1cm,angle=360]{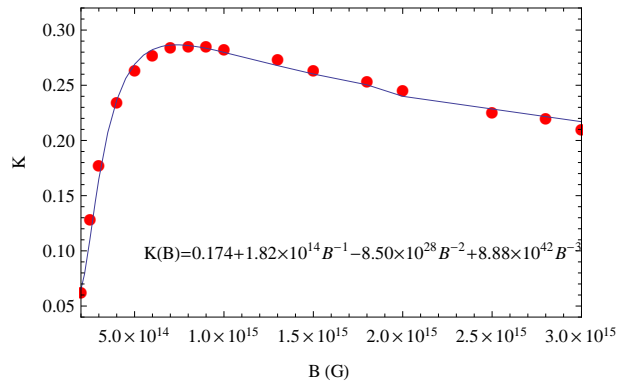}
  \caption{The diagram of $K$  versus $B$}
  \label{fg:Fig.3}
  \end{figure}
Furthermore, the analytic expression of $K$ and $B$ is obtained by fitting the data of Table 1,
\begin{equation}
K(B) = 0.174 + \frac{1.82\times 10^{14}}{B}- \frac{8.50 \times 10^{28}}{B^{2}} + \frac{8.88\times 10^{42}}{B^{3}}, 
\end{equation}
From the definition of $K= \Gamma/\Gamma^{'}$, it is obvious that the value of $K$ is determined by $B$, and is irrelevant to $T$.

Since the updated expression of $\rho_{\rm e}$ is utilized, the values of
$\Gamma$ are universally lower than those of $\Gamma^{'}$
slightly, however, we cannot differentiating Eq.(8) directly. To the contrary,
Eq.(9) is very useful in differential calculations, especially in calculating
magnetic fields' evolutionary timescales. In order to obtain a second-order
differential equation of $B$ and $t$, we may combing Eq.(8) with Eq.(9). By
using Eq.(10), we obtain an approximation relation between $\Gamma$ and $\Gamma^{'}$,
\begin{eqnarray}
&&\Gamma = K(B)\Gamma^{'}= 0.018 K(B)\Lambda (B, T) \nonumber\\
&&\times \int_{Q}^{E_{\rm F}({\rm e})}(E_{\rm e}^{2}-0.261)^{\frac{1}{2}}
E_{\rm e}(E_{\rm e}-Q)^{2}dE_{\rm e}~~.
\end{eqnarray}
\section{Magnetic field decay of magnetars in SNRs}
\subsection{Superhigh magnetic fields and their evolutionary timescales}
In order to investigate the whole process of the decay of superhigh
magnetic fields, to begin with, let us make two approximations:
(1) A magnetar can be treated as a common NS with a total
mass of $M= 2.8 \times 10^{33}$ g (that is about 1.4 times the solar
mass) and a radius of $R= 10^{6}$ cm; (2) The whole electron capture timescale
is equal to the decay timescale of superhigh magnetic fields (without
consideration of the modified Urca process).

As discussed in Sec.1, if the ${}^3P_{\rm 2}$ neutron Cooper pairs are
destroyed by the outgoing EC neutrons, both the anisotropic superfluid
and superhigh magnetic fields produced by the aligned magnetic moments
of the ${}^3P_{\rm 2}$ Cooper pairs will disappear gradually.  Employing
Eq.(10) can allow us to gain a differential equation
\begin{eqnarray}
&&\frac{d\Gamma}{dt} \approx K(B)\frac{d\Gamma^{'}}{dt}= 0.018 K(B)\Lambda(B, T)(E_{\rm F}({\rm e})-Q)^{2}\nonumber\\
&&E_{\rm F}({\rm e})(E^{2}_{\rm F}({\rm e})
-m^{2}_{\rm e}c^{4})^{\frac{1}{2}}43.44 \times\frac{1}{4}B^{-\frac{3}{4}}B^{-\frac{1}{4}}_{\rm cr}\frac{dB}{dt},
\end{eqnarray}
where $\frac{d(0.018K(B)\Lambda(B, T))}{dt}$ is ignored because of its too low
value ($0.018K(B)\Lambda(B, T)\sim 10^{-17}\sim 10^{-18}$(MeV)\\
$^{-5}$~s$^{-1}$, $\frac{d(0.018K(B)\Lambda(B, T)}{dt}\sim \frac{\Delta(0.018K(B)\Lambda(B, T)}{\Delta t}\sim \\
\frac{10^{-17}\sim 10^{-18}{\rm (MeV)^{-5}~s^{-1}}}{10^{13}\rm s}\sim
10^{-30}\sim 10^{-31}$(MeV)$^{-5}$~s$^{-2}$, the integral term $\int_{Q}^{E_{\rm F}
({\rm e})}(E_{\rm e}^{2}-0.261)^{\frac{1}{2}}E_{\rm e}(E_{\rm e}-Q)^{2}dE_{\rm e}\sim
10^{10}-10^{6}$(MeV)$^{5}$, and its time derivative $\sim 10^{-3}-10^{7}$(MeV)$^{5}$~s$^{-1}$
assuming $\Delta t\sim 10^{6}$ yrs). Using binomial expansion theorem, the term
$(E^{2}_{\rm F}({\rm e})-m^{2}_{\rm e}c^{4})^{\frac{1}{2}}$ can be expanded as:
\begin{eqnarray}
&&(E^{2}_{\rm F}(e)-m^{2}_{\rm e}c^{4})^{\frac{1}{2}}= E_{\rm F}(e)(1-m^{2}_{\rm e}c^{4}/2E^{2}_{\rm F}(e)\nonumber\\
&&- m^{4}_{\rm e}c^{8}/8E^{4}_{\rm F}(e)+ \cdots )\approx 43.44(\frac{B}{B_{\rm cr}})^{\frac{1}{4}}\nonumber\\
&& \times (1-542B^{-\frac{1}{2}}-146932B^{-1}+ \cdots)~~.
\end{eqnarray}
Since $542B^{-\frac{1}{2}}\sim 10^{-5}$ and $146932B^{-1}\sim 10^{-10}$,
we will reserve the first term in the bracket of the binomial
expansion in the following calculations. Since a normal radio pulsar
can be treated as a system of magnetic dipoles, there is an
approximation relation of $\mu= \frac{1}{2}B R_{\rm 6}^{3}$, where $
\mu$, $B$ and $R_{\rm 6}$ are the dipole magnetic moment, the dipole
magnetic field strength and the radius of the star in units of
$10^{6}$ cm, respectively \citep{Shapiro83}.  Like normal radio pulsars,
a magnetar can be seen as a magnetic dipole system, the above approximation
relation is also hold in a magnetar.  Assuming that one outgoing EC
neutron can destroy one ${}^3P_{\rm 2}$ Cooper pair (see Appendix A),
the decay rates of magnetic fields of magnetars can be estimated as
\begin{equation}
\frac{dB}{dt} =\frac{2}{R_{\rm 6}^{3}}\frac{d\mu }{dt}=\frac{2}{R_{\rm 6}^{3}}
(-\Gamma 2\mu_{\rm n} n_{\rm e}V({}^3P_{\rm 2}))~~,
\end{equation}
where  $V({}^3P_{\rm 2})$ denotes the volume of the ${}^3P_{\rm 2}$ anisotropic
neutron superfluid, $V({}^3P_{\rm 2})=\frac{4}{3}\pi R_{\rm 5}^{3}$ cm$^{3}$,
$R_{\rm 5}=10^{5}$ cm, and $n_{\rm p}= n_{\rm e}$= 9.6$\times 10^{35}$ cm$^{-3}$
setting $\rho= \rho_{\rm 0}$.  Since $\Gamma$ in this paper represents the
effective electron capture rate, Eq.(14) deviates greatly from Eq.(11) in
Paper 2, though they are exactly like.  Be note, in the interior of a NS, the
processes of EC and $\beta$-decay exist at the same time, which is required by
electric neutrality, the depleted protons/electrons are recycled for many
times, so the alteration of $Y_{\rm p}/Y_{\rm e}$ could be very small. From
Eq.(14), we get
\begin{equation}
\frac{d\Gamma}{dt} =\frac{-R_{6}^{3}}{4 \mu_{\rm n} n_{\rm e}V({}^3P_{2})}\frac{d^{2}B}{dt^{2}}~~, 
\end{equation}
where $\mu_{n}$= 0.966 $\times 10^{-23}$ erg~G$^{-1}$ is the absolute value of
the neutron abnormal magnetic moment.  Combining Eq.(12) with Eq.(15) and
eliminating $\Gamma$ yields a second-order differential equation:
\begin{eqnarray}
&&\frac{d^{2}B}{dt^{2}}+ 4.8002 \times10^{24}(2.60388\times 10^{-14}+~1.98233\nonumber\\
&& \times 10^{-30}B)(0.174+ 1.82\times 10^{14}B^{-1}-8.50\times10^{28}\nonumber\\
&&B^{-2}+ 8.88\times 10^{42}B^{-3})(1.98\times 10^{-25}B^{\frac{1}{4}}-1.394 \nonumber\\
&&\times10^{-21}+2.458\times10^{-18}B^{-\frac{1}{4}})\frac{dB}{dt}=0,
\end{eqnarray}
where~$T_{0}=~2.6\times 10^{8}$~K is used. For the purpose of calculating
the whole electron capture time $t$, we can treat this second-order differential
equation as follows: Firstly, decreasing the order of Eq.(16) gives a first-order
differential equation
\begin{eqnarray}
&&\frac{dB}{dt}=-(-1.2125~\times10^{36}B^{\frac{-9}{4}}+7.73615\times10^{32} \nonumber\\
&& B^{-2}-1.2558\times 10^{29}B^{\frac{-7}{4}}+2.0725\times10^{22}B^{\frac{-5}{4}}- \nonumber\\
&& 1.46924\times10^{19}B^{-1}+2.78251~\times 10^{15}B^{\frac{-3}{4}}-2.1571 \nonumber\\
&& \times10^{8}B^{\frac{-1}{4}}+17.3762B^{\frac{1}{4}}+7.69529\times10^{-8}B^{\frac{3}{4}} \nonumber\\
&& -3.27316\times 10^{-11}B-~30.5838~log_{10}B+~3.7193 \nonumber\\
&&-1.15403\times\times 10^{-15}B^{\frac{5}{4}}+2.32557\times 10^{-24}B^{\frac{7}{4}} \nonumber\\
&& 10^{-27}B^{2}+1.45703\times 10^{-31}B^{\frac{9}{4}})+C.
\end{eqnarray}
Secondly, inserting the boundary condition: $dB/dt= 0$ when $B=B_{\rm th}$,
into Eq.(17) determines the constant of integral $C=-992890.0$; Thirdly, integrating
over $B$ gives a general expression of $t$
\begin{eqnarray}
&&t~=\int _{B_{\rm i}}^{B_{\rm f}}-(-1.2125\times10^{36}B^{\frac{-9}{4}}+7.73615\times10^{32} \nonumber\\
&&B^{-2}-1.2558\times 10^{29}B^{\frac{-7}{4}}+2.0725\times10^{22}B^{\frac{-5}{4}}-    \nonumber\\
&& 1.46924\times 10^{19}B^{-1}+2.78251\times 10^{15}B^{\frac{-3}{4}}-2.1571 \nonumber\\
&&\times10^{8}B^{\frac{-1}{4}} +17.3762B^{\frac{1}{4}}+~.69529\times 10^{-8}B^{\frac{3}{4}}\nonumber\\
&& -3.27316\times10^{-11}B-30.5838~log_{10}B+3.7193\times\nonumber\\
&&10^{-15}B^{\frac{5}{4}}+2.32557\times10^{-24}B^{\frac{7}{4}}-1.15403\times10^{-27}  \nonumber\\
&&B^{2}+1.45703\times 10^{-31}B^{\frac{9}{4}}~+~992890.0)^{-1}dB. 
\end{eqnarray}
Finally, using integral transform $B^{\frac{1}{4}}\rightarrow x$ and
$dB \rightarrow 4x^{3}dx$ gives the final expression of $t$

\begin{eqnarray}
&&t=\int_{B^{0.25}_{\bf f}}^{B^{0.25}_{\bf i}}(-1.2125\times10^{36}x^{-9}+7.73615x^{-8}-\nonumber\\
&&1.2558\times10^{29}x^{-7}+2.0725\times 10^{22}x^{-5}-1.46924  \nonumber\\
&&\times 10^{19}x^{4}+~2.78251~\times 10^{15}x^{-3}-2.1571\nonumber\\
&&\times10^{8}x^{-1}+17.3762x+~7.69529~\times 10^{-8}x^{3}\nonumber\\
&&-3.27316\times 10^{-11}x^{4}-30.5838~log_{10}x^{4}+3.7193\times \nonumber\\
&&10^{-15}x^{5}~+~2.32557~\times 10^{-24}x^{7}-1.15403\times10^{-27} \nonumber\\
&&x^{8}+1.45703\times 10^{-31}x^{9}+992890.0)^{-1}4x^{3}dx,
\end{eqnarray}
where $x\geq B_{\rm th}^{0.25}~=~3524.05$.
In order to investigate the characteristics of Eq.(19), we introduce a
variable $F(x)$ to denote the integrated function, $F(x)=(-1.2125\times
10^{36}x^{-9}+7.73615\times 10^{32}x^{-8}-~1.2558~\times 10^{29}x^{-7}+
2.0725\times 10^{22}x^{-5}-1.46924\times 10^{19}x^{-4}+~2.78251\times10^{15}
x^{-3}-2.1571\times 10^{8}x^{-1}+~17.3762x+7.69529~\times 10^{-8}
x^{3}-3.27316\times 10^{-11}x^{4}-30.5838~log_{10}x^{4}~+~3.7193\times 10^{-15}
x^{5}+~2.32557\times 10^{-24}x^{7}-~1.15403\times 10^{-27}x^{8}+1.45703
\times 10^{-31}x^{9}+992890.0)^{-1}4x^{3}$, and make a schematic diagram of
$F(x)$ as a function of $x$, shown as in Fig.4.
\begin{figure}[th]
\centering
  \vspace{0.5cm}
  \includegraphics[width=8.1cm,angle=360]{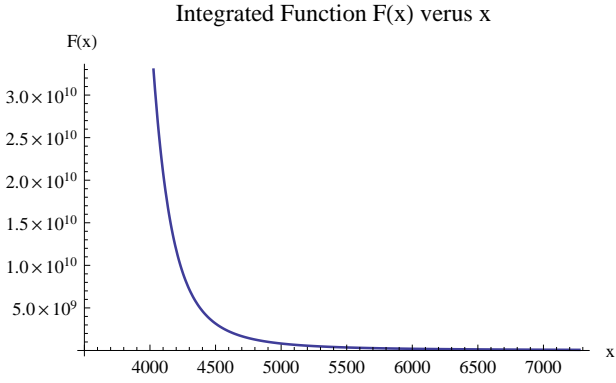}
  \caption{The diagram of $F(x)$ versus $x$}
  \label{fg:Fig.4}
  \end{figure}
From Fig.4, in the integral interval of $[(x_{1}\sim x_{2}),~ x_{1}~\leq~x_{2}]$, the
function~$F(x)$~is convergent, and hence is integrable, where $x =~3524.05$ is a
singularity of $F(x)$.  By solving Eq.(19), we gain the whole electron capture
time (or the superhigh magnetic field's decay timescale), $t\approx 9.2947 \times
10^{13}$ s = $2.9507 \times 10^{6}$ yrs when $B_{\rm i} = 3.0 \times 10^{15}$ G and
$B_{\rm f}= B_{\rm th}$; in the same way, if $B_{\rm i}= 3.0 \times 10^{15}$ G and $B_
{\rm f} = 4.0 \times 10^{14}$~G, $t\approx 1.4737\times 10^{12}$ s =4.6785 $\times
10^{4}$ yrs, corresponding to $L_{\rm X}\sim(10^{37}\sim 10^{34})$ erg~s$^{-1}$.
Furthermore, the fitting curves of Log$_{10}B$ versus Log$_{10}t$ for different magnetic
field ranges are shown in Fig.5.
\begin{figure}[th]
\centering
  \vspace{0.5cm}
  \includegraphics[width=8.1cm,angle=360]{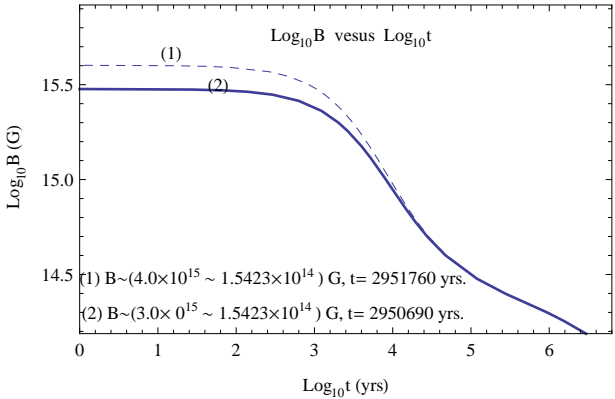}
  \caption{Superhigh magnetic field $B$ as a function of time $t$.}
  \label{fg:Fig.5}
  \end{figure}

From Fig.5, the magnetic field decreases with increasing time obviously. Customarily,
$B= 3.0\times 10^{15}$ G is assumed to be the possible initial magnetic field, while
$B= 4.0 \times 10^{15}$ G is assumed to be the upper limit of magnetic field of a
magnetar by some authors (e.g., recent papers of Peng \& Tong 2007, 2009, and
Gao et al 2011). However, the difference between the evolution timescales of these
two fields is only $\Delta t\simeq 1070$~yrs, obtained from Eq.(19) (as shown in Fig.5), which
implies a substantial positive correlation between the magnetic field and its
decay rate.

Since the initial value of internal temperature $T$ is far higher than
its current value for a magnetar, we arbitrarily assume $T_{0}\sim 2.75\times 10^{8}-
1.75\times 10^{8}$~K within a permitted and plausible temperature range under our
magnetar model. Repeating all the calculations above gives the relation of $t$ and $T_{0}$, shown as in Fig.6.

\begin{figure}[th]
\centering
  \vspace{0.5cm}
  \includegraphics[width=8.1cm,angle=360]{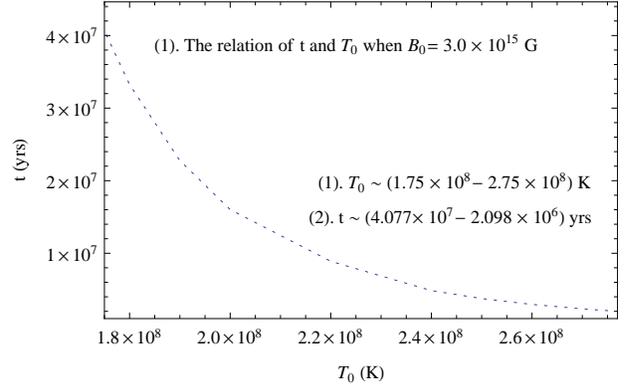}
  \caption{The magnetic field evolution timescale $t$ as a function of time $T_{0}$.}
  \label{fg:Fig.6}
  \end{figure}
From Fig.6, it is apparent that the lower $T_{0}$, the higher $t$ if $B_{0}$ is invariable.
When $B_{0}=3.0\times 10^{15}$ G, $T_{0}\sim 2.75\times 10^{8}-1.75\times 10^{8}$~K, then
$t\sim 2.098\times 10^{6}-~4.077\times 10^{7}$~yrs.
\subsection{The initial internal fields of magnetar progenitors}
Although we have presented a reasonable explanation for the origin of superhigh
magnetic fields in the previous work \citep{Peng07, Peng09}, many issues concerning
magnetars remain unsolved.  So far, the magnetic fields of magnetars obtained from
the observations are just their surface dipolar magnetic fields $B_{\rm dip}$,
assuming a simple magnetic dipole spin-down model. What is the relationship between
the surface dipolar magnetic field and the internal magnetic field in a magnetar?
How strong is the initial internal magnetic field $B_{\rm i}$ of a magnetar
progenitor? How long will such an intense field ($B_{\rm i}$) continue to decay?
What will eventually happen in the interior of a magnetar when $B$ drops
below $B_{\rm th}$? All these questions are all very basic, and remain open.

In this part, motivated by SNR associations, we try to carry out the studies of
$B_{\rm i}$ for magnetar progenitors. Observations indicate that SNRs have been
expanding, and interacting with their surroundings since the supernova explosions.
Therefore, the ages of SNRs may be computed by modeling their morphologies at the
current epoch. The true ages of magnetars obtained from the ages of their SNRs are
independent of the stars' properties, and thus basically unbiased even if AXPs or
SGRs are strange stars \citep{Zhang00, Xu06}. Table 2 shows the data on 9 claimed
magnetar-supernova remnant associations, which are cited from McGillAXP/SGR online
catalog updated except for SGR 1806-20 and SGR 1900+14.
\begin{table*}[t]
\small
\caption{The calculated values of the initial internal magnetic fields of magnetar progenitors  \label{tb1-2}}
\begin{tabular}{@{}crrrrrrr@{}}
\tableline
Source& $P$& $\dot{P}$& $B_{\rm dip}$ & SNR & $t_{\rm SNR}$ &$B_{\rm i}^{a}$& Ref$^{d}$\\
\tableline
SGR 0526-66             &8.0544    &3.8     &5.6          &N49                      &5.0             &5.06          &[1, 2] \\
SGR 1806-20             &7.6022   &75      &24           &G10.0-0.3$^{\ddag}$      &$<10^{\ddag}$   &$\leq 29.3$  &[3, 4] \\
SGR 1627-41             &2.5946    &1.9     &2.20       &G33.70-01                &5.0             &2.204          &[5, 6] \\
SGR 1900+14            &5.1998   &9.2     &7.0       &G42.8+0.6                &$<10^{\S}$  &$<13.71$        &[7, 8] \\
1E 2259+586             &6.9789   &0.048   &0.59         &CTP109                 &$\sim 10$       & $\sim 1.5435$         &[9, 10] \\
1E 1841-045             &11.7829   &3.93    &6.9        &Kes73                    &2               &7.497         &[11, 12] \\
1E 1547.0-5408          &2.318    &2.318   &2.2        &G327.24-0.13             &$<1.4$          &$<2.202$           &[13, 14] \\
CXOU J171405.7$^{\dag}$ &3.82535   &6.40    &5.0    &CTB37B                   &4.9             &5.509         &[15-17] \\
AX J1845-0258$^{\dag}$  &6.97127   &No      &No       &G29.6+0.1                &$<8$            &$\leq 40^{b}$  &[18, 19]\\
 \tableline
  \end{tabular}
\tablecomments{The units of period $P$, period derivative $\dot{P}$, the surface dipolar magnetic field $B_{\rm dip}$,
SNR's age $t_{\rm SNR}$ and the initial internal magnetic field $B_{\rm i}$ are s, $10^{-11}$s s$^{-1}$,
$10^{14}$ G, $10^{3}$ yrs and $10^{14}$ G, respectively. \\
$^{a}$ The values of the initial internal magnetic fields of magnetar progenitors $B_{\rm i}$ are gained by using Eq.(18) and $t \simeq t_{\rm SNR}$.\\
$^{b}$ Since AXP 1E 2259+586 could be associated with accretion (see Paper 4), and its present value of $B_{\rm dip}$ is far
less than $B_{\rm th}$, its maximum value of $B_{\rm i}$ has to be estimated under our model.  \\
$^{c}$ Since some important parameters (eg., $\dot{P}$, $B_{\rm dip}$, the soft X-ray luminosity $L_{\rm X}$, and so on) of magnetar candidate AX J1845-0258 are uncertain, its maximum value of $B_{\rm i}$ has to be estimated under our model.\\
$^{d}$ 1$-$\citep{Kulkarni03}; 2$-$\citep{Klose04}; 3$-$\citep{Kulkarni93}; 4$-$\citep{Marsden01};
5$-$\citep{Corbel99}; 6$-$\citep{Wachter04}; 7$-$\citep{Hurley99}; 8$-$\citep{Mazets99}
9$-$\citep{Green89}; 10$-$\citep{Rho97}; 11$-$\citep{ Sanbonmatsu92}; 12$-$\citep{Vasisht97};
13$-$\citep{Camilo07};  14$-$\citep{Gelfand07}; 15$-$\citep{Aharonian08};
16$-$\citep{Halpern10}; 17$-$\citep{Horvath11}; 18$-$\citep{Gaensler99}; 19$-$\citep{Vasisht00}.\\
$^{\dag}$ This candidate is unconfirmed. \\
$^{\ddag}$ Cited from Kulkani \& Frail(1993) and Marsden et~al.(2001).
$^{\S}$ Cited from Hurley et al.(1999). All primitive data are from McGillAXP/SGR online catalog of 2 March 2012(http://www.physics.
mcgill.ca/$^{\sim}$pulsar/magnetar/ main.html) and references cataloged.}
\end{table*}
As is known to us, all the known SNRs associated with common radio pulsars are very
young, $t_{\rm SNR}\ll 10^{6}$ yrs.  From Table 2, the ages of all the SNRs are not
more than 10,000 yrs, which implies that the associated magnetars are more younger,
compared with common radio pulsars.  Perhaps these magnetars born with different
physical properties (e.g., the equations of state, magnetic fields, inner temperatures,
and so on) have experienced evolutionary routes that differ from those of common radio
pulsars. For the sake of computing conveniently, we assume a simple magnetic dipole
spin-down model, then the current magnetic field inside a magnetar is about its
surface dipole field, i.e., $B_{\rm f}\simeq B_{\rm dip}$ under this assumption. The
field decay timescale of a magnetar equals approximately to the age of its SNR, i.e.,
$t \simeq t_{\rm SNR}$, base on the results of ZX2011.  Combining Eq.(19) with Table 2
gives the values of $B_{\rm i}$ for magnetar progenitors. The calculated results show
that the values of $B_{\rm i}$ are concentrated primarily between $2.0 \times 10^{14}$ G
and $2.93 \times 10^{15}$ G when $T_{0}\sim 2.60\times 10^{8}$ K. If $T_{0}\sim 2.75\times
10^{8}- 1.75\times 10^{8}$~K, there still be $B_{\rm i}\sim(10^{14}-10^{15})$~G, not
exceeding the upper limit of magnetic field of $4.0\times 10^{15}$ G. The calculated results
of $B_{\rm i}$ for magnetar progenitors illustrate that our magnetar model is consistent
theoretically.

\subsection{ The spin-down ages of magnetars and the ages of their SNRs}
As pointed out in Sec.1, the NS's spin-down age, also called the characteristic
age, can be generally expressed as, $t_{\rm Spin} = P/(n-1)\dot{P}$.  In all kinds
of catalog on pulsars, the spin-down ages of pulsars are usually evaluated by a
simple magnetic dipole spin-down model, $t_{\rm Spin}=P/2\dot{P}$ ($n = 3$). In
ZX2011, authors found that there is a strong and significant positive correlation
between Log$_{10}$(SNR(Age)/Spindown(Age)) and Log$_{10}B$ via statistical
analysis of radio pulsars associated with SNRs. They argued that, as NSs get older,
their spin periods become longer duo to their spin-down torques, the decay of
magnetic fields cause $\dot{P}$ to be far less than the mean values of $\dot{P}$
in history, their characteristic ages, inferred from parameters $P$ and $\dot{P}$
at the current epoch, will be larger than their real ages, denoted as $t_{\rm Real}$.
In a word, it's the dipolar magnetic field decay that plays a significant role
in making a NS look older than it really is.

In this part, we investigate the spin-down ages of magnetars and the ages
of their SNRs, and made a diagram of Log$_{10}$(SNR(Age)/Spindown(Age)) versus
Log$_{10}B$.
\begin{figure}[th]
\centering
  \vspace{0.5cm}
  \includegraphics[width=8.1cm,angle=360]{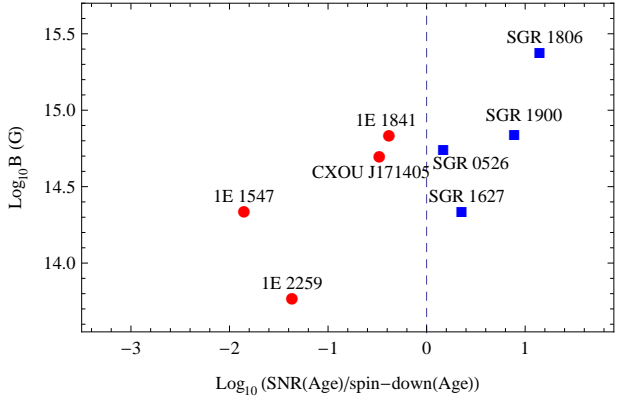}
  \caption{Log$_{10}$(SNR(Age)/Spindown(Age)) versus Log$_{10}B$. The range of
$B$ is $(1.5423 \times 10^{14}\sim 3.0 \times 10^{15})$~G. The dashed line
corresponds to $t_{\rm SNR}= t_{\rm Spin}$.}
  \label{fg:Fig.7}
  \end{figure}

From Fig.7, an obvious correlation has been proved between Log$_{10}B$ and
Log$_{10}$(SNR(Age)/Spindown(Age) for magnetars associated with SNRs. It is
worthwhile to note that all AXPs associated with SNRs are on the left of the
dashed line whereas all SGRs associated with SNRs are on the left of the line.
The causes of this will be discussed at length in the following.

Since a magnetar can be seen as a magnetic dipole system, the above suggestion
that `the dipolar magnetic field decay plays a significant role in making a NS
look older' is also applicable to magnetars.  For all the  AXPs (including
candidates) with dipole magnetic fields $B< 7.0 \times 10^{14}$ G,
there have been no super-bursts or giant flares, at which huge energies ($\sim
10^{43}-10^{47}$ erg) are suddenly released.  This suggests that all AXPs could have
experienced relatively `normal' decay of their dipole magnetic fields compared
with SGRs associated with SNRs, and thus have lower braking indexes, $n< 3$.

In previous works, many authors proposed various models to explain why the
observed braking index $n < 3$, eg., neutrino and photon radiation coming from
superfluid neutrons may brake the pulsars \citep{Peng82}; both magnetic dipole
radiation and the propeller torque applied by the debris disk may cause spin-down
of pulsars \citep{Alpar01, Menou01}; the combination of dipole radiation and the
unipolar generator may cause $n$ decrease greatly \citep{Xu01,Wu03}; a variation
of the torque function is important attribution for low braking index \citep{Allen97};
additional torques due to accretion may cause $n$ decrease\citep{Menou01, Chen06};
$n < 3$ may be due to the decay of magnetic field strengthes\citep{Blandford88,
Lin04, Chen06, Zhang11}, and so on. As we know, magnetars are high magnetized NSs.
They universally possess very strong surface dipole magnetic fields ($\sim 10^{14}
-10^{15}$ G) with rare exceptions (eg., SGR 0418+5729, 1E 2259+586 and unconfirmed
candidate Swift J1822.3-1606). Thus, we favor the braking model with changing
magnetic field strengthes, ie., the decay of magnetic field leads to $n< 3$ for
magnetars. In our magnetar model, the lower braking indices $n$ ($n< 3$) of AXPs are
supposed to be correlated with the dipole magnetic fields $B$ and their decay rates
$dB/dt$.  In order to validate this assumption,, we investigate the phenomenon of
$n< 3$ for neutron stars (including common radio pulsars and magnetars) under pure
magnetic dipole spin-down model, theoretically.

As we know, the spin frequency $\Omega$ of pulsars decreases with time, and the time
derivative of $\Omega$ is proportional to some power of $\Omega$,
\begin{equation}
I\dot{\Omega} =~-K~\Omega^{n}~, 
\end{equation}
where $K=2B^{2}R^{6}sin^{2}\theta~/3c^{2}$, $\theta$ is the inclination of the magnetic
axis with respect to the rotation axis;  $B$, $R$ $I$ are the surface magnetic field
strength, the radius, and the momentum of inertia of the pulsar, respectively; $c$ is the
velocity of light\citep{Manchester77}. The braking index $n$ of a pulsar can be a
measured by differentiating Eq.(20),
\begin{equation}
n =~\frac{\ddot{\Omega}\Omega}{\dot{\Omega}^{2}}~, 
\end{equation}
where $\ddot{\Omega}$ is the second order time derivative of $\Omega$. In the model that
assumes spin-down is due to pure magnetodipole radiation with a constant magnetic field,
we obtain the ideal values of $n$ in Eqs.(20-21), $n=3$ \citep{Manchester77, Blandford88,
Menou01, Chen06, Zhang11}.  As mentioned in Sec.1, the observed values of $n$ from Eq.(21)
always deviate from 3 expected for pure magnetodipole radiation model (only in this case,
$t_{\rm Spin}=t_{\rm SNR}$).  With respect to the case of $n<3$, the main and possible
causes have been  provided, as listed above. In this paper, we will simply discuss the error(or deviation)
of $n$ caused by the derivation of Eq.(21) itself.  From the deduction above, Eq.(21) is obtained
by differentiating Eq.(20), assuming $I$ and $K$ are constant. As a result, $n$ is mainly determined
by $\Omega$ (or spin period $P$, $P=2\pi/\Omega$) and its time derivatives, and is irrelevant to
the other quantities. Actually, it's possible that the quantities of $B$, $\theta$, $I$, $R$ and
$\Omega$ change in varying degrees, and changes of $B$, $\theta$, $I$ and/or $R$ with $\Omega$
will induce a braking index $n\neq 3 $.  Hence, a modification of Eq.(21) is necessary.
Keeping the quantities of $\theta$, $I$ and $R$ unchanged (the variances of these quantities
are small usually), we re-differentiate Eq.(20) under pure magnetic dipole spin-down model, and get
\begin{equation}
\ddot{\Omega}=~\frac{d\dot{\Omega}}{dt}=\frac{-2R^{6}sin^{2}\theta}{3c^{2}I}(2B\dot{B}\Omega^{3}
+~3B^{2}\Omega^{2}\dot{\Omega}) ~, 
\end{equation}
Inserting Eq.(20) and Eq.(22) into Eq.(21), we have
\begin{eqnarray}
&&n=~\frac{-2R^{6}sin^{2}\theta}{3c^{2}I}(\frac{2B\dot{B}\Omega^{4}} {\dot{\Omega}^{2}}
+~\frac{3B^{2}\Omega^{3}\dot{\Omega}}{\dot{\Omega}^{2}}) \nonumber\\
&&=\frac{-2R^{6}sin^{2}\theta}{3c^{2}I}(\frac{9c^{4}I^{2}\dot{B}}{2R^{12}sin^{4}\theta~\Omega^{2}B^{3}}
-~\frac{9c^{2}I}{2R^{6}sin^{2}\theta~}) \nonumber\\
&&=3-\frac{3c^{2}I \dot{B}}{R^{6}sin^{2}\theta~\Omega^{2}B^{3}}=
3-\frac{3c^{2}P^{2}I}{4\pi^{2}R^{6}sin^{2}\theta~}\frac{\dot{B}}{B^{3}}\nonumber\\
&&=~3-~\frac{3c^{2}P^{2}I}{4\pi^{2}R^{6}sin^{2}\theta~}\frac{|dB/dt|}{B^{3}},
\end{eqnarray}
where $\dot{B}$ denotes the absolute value of decay rate of dipole magnetic field without
considering any other torque (ie., the dipole magnetic field normally decays). We can estimate the
magnitude of the second term in Eq.(23) as following: for a common radio pulsar,
$I\sim 10^{45}$~g~cm$^{2}$, $sin^{2}\theta \sim 1$, $R\sim 10^{6}$~cm, $P\sim 1$~s,
$c=3.0\times 10^{10}$~m~s$^{-1}$, $\frac{\dot{B}}{B^{3}}\sim \frac{|dB/dt|}{B^{3}}\sim
\frac{|\Delta B/\Delta t|}{B^{3}}\sim \frac{1}{B^{2}t}\sim 10^{-38}-10^{-39}$~G$^{-2}$
~s$^{-1}$ ($\Delta B\sim B\sim 10^{12}$~G, $\Delta t \sim t\sim 10^{14}-10^{15}$~s),
$\frac{3c^{2}P^{2}I}{4\pi^{2}R^{6}sin^{2}\theta~}\frac{\dot{B}}{B^{3}}\sim 10^{-8}-10^{-9}$;
for a canonic magnetar, $P\sim 10$~s, $\frac{\dot{B}}{B^{3}} \sim \frac{1}{B^{2}t}\sim
10^{-41}-10^{-43}$~G$^{-2}$~s$^{-1}$ ($\Delta B\sim B\sim 10^{14}-10^{15}$, $\Delta t \sim
t\sim 10^{13}$~s), $\frac{3c^{2}P^{2}I}{4\pi^{2}R^{6}sin^{2}\theta~}\frac{\dot{B}}{B^{3}}
\sim 10^{-9}-10^{-10}$, therefore, when all the quantities of $\theta$, $I$, $R$ and $P$
(or $\Omega$) remain unchanged, and the dipole magnetic field normally decays, $n\approx 3$,
in the ideal situation of $\dot{B}=0$, $n=3$. However, if there are substantial changes of
$\theta$, $I$, $R$ and $P$ (at least one quanity varies), the effects of $\frac{\dot{B}}{B^{3}}$
on $n$ cannot be ignored. Under pure magnetic dipole spin-down model, we produce the diagrams
of $dB/dt-B$ and $\frac{\dot{B}}{B^{3}}-B$ by using the method of curve fitting, shown as in
Fig.8.

\begin{figure*}[htb]
\begin{center}
\begin{tabular}{cc}
\scalebox{0.89}{\includegraphics{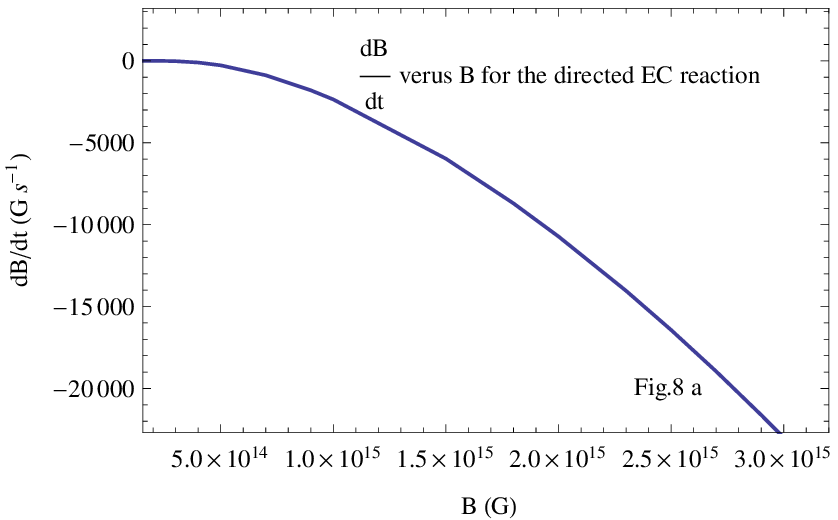}}&\scalebox{0.89}{\includegraphics{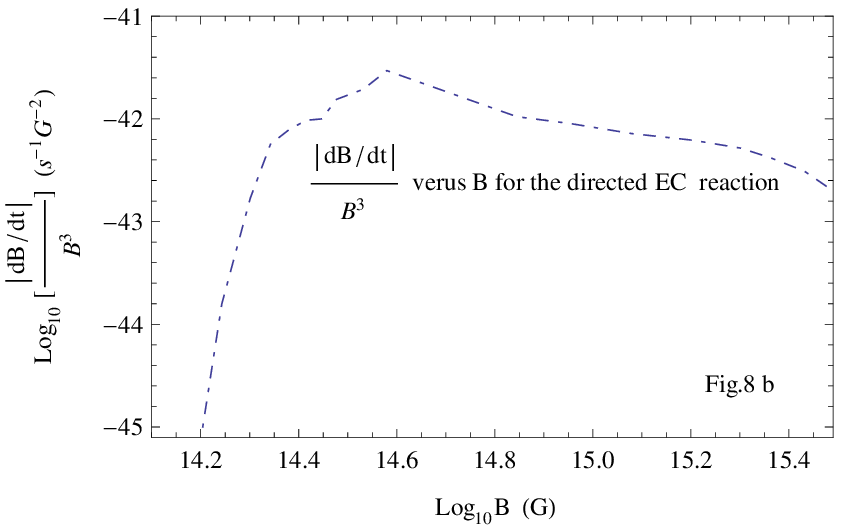}}\\
(a)&(b)\\
\scalebox{0.89}{\includegraphics{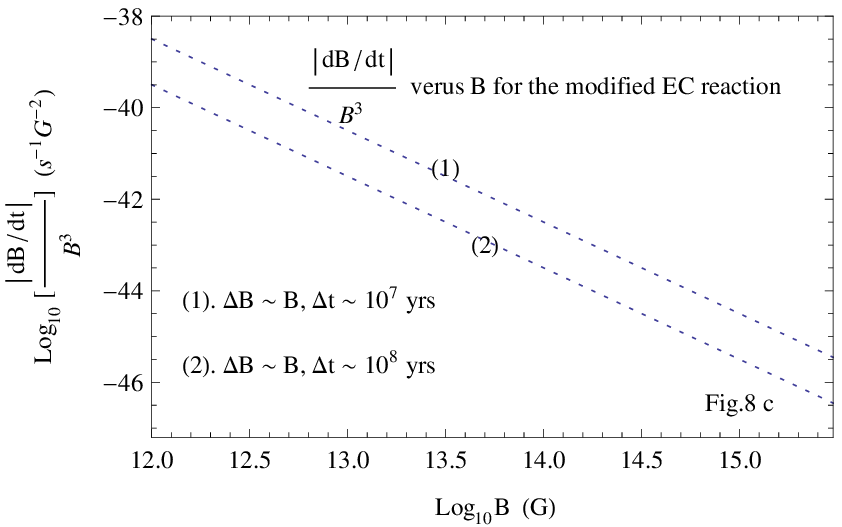}}&\scalebox{0.89}{\includegraphics{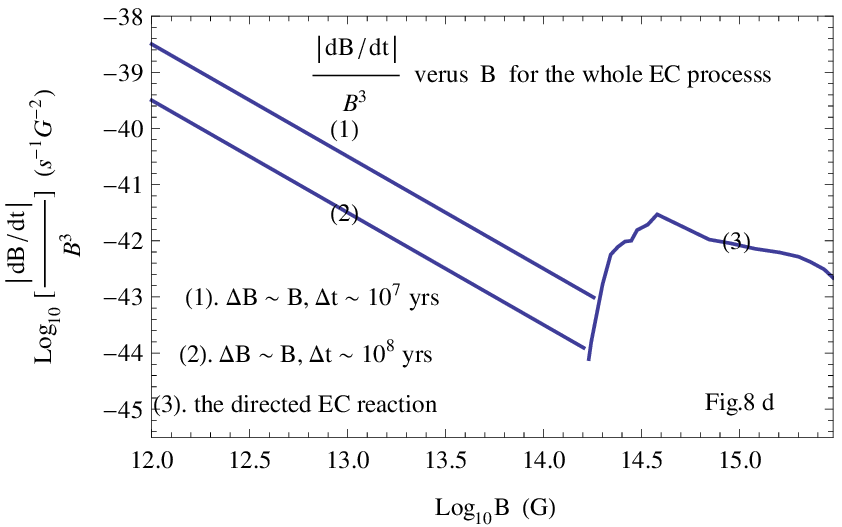}}\\
(c)&(d)\\
\end{tabular}
\end{center}
\caption{The diagrams of $dB/dt-B$ and $\frac{\dot{B}}{B^{3}}-B$
 \label{fg:Fig.8}}
\end{figure*}

Fig.8 is composed of four sub-figures. The fitted curves in Fig.8 a and Fig.8 b are
obtained from Eq.(17). Fig.8 $a$ shows that $\frac{dB}{dt}$ decreases with decreasing
$B$ significantly in the directed EC process. In Fig.8 $b$, $\frac{\dot{B}}{B^{3}}$
increases with decreasing $B$ when $B\sim 3.0\times 10^{15}\sim(4-5)\times 10^{14}$ G,
because $B^{3}$ decreases faster than $\dot{B}$, whereas $\frac{\dot{B}}{B^{3}}$
decreases with decreasing $B$ when $B\sim(4-5)\times 10^{14}\sim 1.5423\times 10^{14}$ G,
and $\frac{\dot{B}}{B^{3}}=0$ when $B=B_{\rm th}$ because the directed EC reaction ceases.
It's worth noting that the modified EC reaction always proceeds in the interior of a neutron star with
any magnetic field strength, if the directed EC reaction exists, the modified EC reaction
can be ignored (See Paper 1 and Paper 2). If the modified EC reaction dominates, the total
magnetic decay rate $\frac{dB}{dt}$ (or $\dot{B}$) may be determined by many other factors,
eg., Ohmic decay, ambipolar diffusion and Hall drift \citep{Goldreich92, Rheinhardt03,
Pons06}. However, we mainly focus on the relation of $\frac{\dot{B}}{B^{3}}$ with $B$, rather
than a specified way of magnetic field when the modified EC reaction dominates. Using
the method of dimensional analysis, we produce the schematic diagrams of $\frac{\dot{B}}{B^{3}}$
as a function $B$ if the magnetic field decay timescale $t$ is given, shown as in Fig.8 $c$.
The curves in Fig.8 $d$ are produced by the superposition of the curves in Fig.8 $b$ and Fig.8
$c$ in a wide range of $B\sim 3.0\times 10^{15}-10^{12}$ G. From Fig.8 $d$, the total change
trend of $\frac{\dot{B}}{B^{3}}$ is that $\frac{\dot{B}}{B^{3}}$ increases with decreasing $B$.
Observations show that for most pulsars, their dipole magnetic fields decay slowly during their
lifetimes, and their observed  braking indices $n$ are smaller than 3.
For young pulsars including AXPs, an obvious correlation has been proved between $n < 3$
and the dipole magnetic fields $B$ as well as their decay rates $dB/dt$, which can be
easily seen from the combination of Fig.8 with Eq.(23).

Among the known 12 AXPs (9 confirmed, 3 candidates), 1E 2259+586
has the weakest dipole magnetic field, $B= 5.9 \times 10^{14}$ G, the shortest
period derivative, $\dot{P}= 4.843 \times 10^{-13}$~s~s$^{-1}$, and the longest
spin-down or characteristic age, $t_{\rm Spin}$ =230 kyr.  All of these date
suggests that the value of $n$ of 1E 2259+586 is less than the ideal value of
$n= 3$.  For 1E 2259+586, its soft X-ray emission could be associated with accretion
\citep{White84, van95}, which is beyond of our model, and the direct EC reaction
ceases due to the weaker field ($B\ll B_{\rm th}$), however the modified EC
reaction still occurs, from which weaker X-ray and weaker neutrino flux are
produced. For this source, the weakest dipole magnetic field $B$ and
super-low rates of decay of $B$ via the modified EC reaction may be the major
causes that contribute to $n< 3$.

In the above  parts, we explain why all AXPs associated with SNRs appear
older than they are.  The reason why all SGRs associated with SNRs appear younger
than their real ages is studied in the follows. Unlike AXPs, all SGRs can emit
short bursts in the hard X-ray/soft gamma-ray range with $E\sim 10^{39}-10^{41}$
erg \citep{Mereghetti08}.  Furthermore, giant flares and intermediate flares (or
giant outbursts) were detected in SGRs associated with SNRs.  Table 3 reports the
details of giant/intermediate flares from these four SGRs associated with SNRs.
\begin{table*}[t]
\small
\caption{Giant/Intermediate flares from four SGRs\label{tb1-3}.}
\begin{tabular}{@{}crrrr@{}}
\tableline
Source      & SGR$^{a}$0526-66 & SGR$^{a}$1900+14  & SGR$^{a}$1806-20 & SGR$^{b}$1627-41 \\
\tableline
Date        &March 5, 1979    &August 27, 1998      & December 27,~2004   & June 17,~1998 \\
Assumed Distance(kpc) &55   &15              &15               & 5.8         \\
Peak Luminosity(erg/s) &$3.6\times 10^{44}$    &$>~1.5\times 10^{44}$ & $(1.6\sim 5)\times 10^{46}$ & $8\times 10^{42}$ \\
Isotropic Energy(erg) &$1.6\times 10^{44}$    &$>~8.3\times 10^{44}$ & $(2\sim 5)\times 10^{47}$    & $8\times 10^{43}$ \\
\tableline
\end{tabular}
  \tablecomments{The sign `a' denotes: The date are cited from \citep{Mereghetti08}.
  The sign `b' denotes: The date are cited from \citep{Mazets99}.}
\end{table*}
In addition, other intermediate flares occurred in SGR 1900+14 on August 29, 1998
\citep{Ibrahim01}, and on April 28, 2001 \citep{Lenters03}. These two intermediate
flares are no longer listed in Table 3 because of their relatively lower energies
$\sim 10^{39}-10^{41}$ erg. The data in Table 3 implies a significant correlation
between giant/intermediate flares and $t_{\rm SNR}> t_{\rm Spin}$ for SGRs. The
decay of diploe magnetic fields and giant/intermediate flares can conjointly affect
spinning behaviors of SGRs associated with SNRs.  However, we suppose that giant/
intermediate flares make SGRs look younger than they are.  Now, an explanation of
$t_{\rm SNR}> t_{\rm Spin}$ for SGRs in the context of the star-quake model of
 magnetars \citep{Thompson02} is presented. The details are as follows:

Giant flares/bursts could be motivated by a large-scale fracture of the crust, driven
by magnetic stresses; the sudden crust's cracking sets the whole magnetar `quaking',
which results in a significant change in configuration of the dipole magnetic field
(including magnetic field strength, magnetic field decay rate, the angle between
magnetic axis and spin axis, moment of inertia, magnetic moment and so on). Such a
change in configuration of the dipole magnetic field could give rise to
unusual increases in spin-down torque as well as braking index $n$.  Therefore,
$\dot{P}$ also increases substantially, which can be illustrated by significant
jumps in the period evolution of SGR 1900+14 after the 27 August 1998 giant flare
\citep{Marsden99, Mereghetti00}. In short, for a SGR associated with a SNR, it is
this change in configuration of the dipole magnetic field that could
produce a significant deviation of $n$ ($n> 3$) and cause the current value of
$\dot{P}$ to be far larger than its mean value in history, the spin-down age will
be far less than its real age. We can take SGR 1806-20 as an excellent example.
From Table 3, the highest energy $E\sim (2-5)\times 10^{47}$ erg was released
during the giant flare of December 27, 2004 from SGR 1806-20 which exceeded all
previous giant/intermediate flares of SGRs. Among the known 23 magnetar, SGR 1806-20
has the strongest dipole magnetic field, $B= 2.4 \times 10^{15}$ G, the largest period
derivative, $\dot{P}=7.5 \times 10^{-10}$ s s$^{-1}$, and the shortest spin-down
or characteristic age, $t_{\rm Spin}$=0.16 kyrs. All of this implies that the value
of braking index of SGR 1806-20 is larger than the ideal value of $n=3$.

An alternative explanation that all SGRs associated with SNRs appear younger
than their real ages is provided in the follows. Glitches (sudden frequency jumps of
a magnitude $\Delta \Omega/\Omega \sim 10^{9}$ to $10^{6}$, accompanied by the jumps
of spin-down rates with a magnitude of $\Delta \dot{\Omega}/\dot{\Omega}\sim 10^{-3}
-10^{-2}$)are common phenomenon in pulsars. After each glitch, a permanent increase in the
pulsar¡¯s spin-down rate usually happens, which causes a slow increase in the pulsar¡¯s surface polar magnetic
field (be note, neither the period nor the spin-down rate is
completely recovered although there is a relaxation after a glitch).  Consequently, some
radio pulsars with many active glitches may evolve into magnetars \citep{Lin04, Chen06}.
As we know, magnetars show many similarities with typical radio pulsars, including the
properties of glitches. The amplitudes of glitches of SGRs are far larger than those of
AXPs and radio pulsars \citep{Mereghetti08}.  The glitches of SGRs associated with SNRs
could be triggered  by stars' `quaking', contributing to magnetars' giant/intermediate
flares \citep{Mereghetti08}.  The significant changes of spin-down rates and dipole
magnetic field strengthes before and after giant flares of SGRs (0526-66, 1806-20,
1900+14 and 1627-41) associated with SNRs are good indications of huge glitches happened
in four SGRs though some of these huge glitches are `missed'(not reported) \citep{Pons12}.
Therefore, a SGR's present spin-down rate may be much higher than its initial value, and
its characteristic age may be shorter than its true age. With respect to AXPs with SNRs,
on one hand, their dipole magnetic fields may also increase to a certain degree via
glitches, on the other hand, the dipole magnetic fields rapidly decay via EC reaction,
but the rates of increase are far less than the rates of decay. Thus, it's the dipolar
magnetic field decay that plays an important role in making an AXP look older than it
really is. It is worth noting that when a glitch happens, Eq.(23) no longer applies because
the quantities of $\theta$, $I$, $R$, $\Omega$(or $P$) may vary to some extent, apart from an increases in $B$.

In ZX2011, authors suggested that the characteristic age of a NS is not available to
estimate its real age, and the physically meaningful criterion to estimate $t_{\rm Real}$
is the NS-SNR association. Their suggestions are in the same way applicable to magnetars
associated with SNRs, according to our analysis above.

\section{Conclusions}
In this paper, based on our modified model, we carry out a study of the
magnetic field decay of magnetars in SNRs.  The main conclusions are as follows:

1.~In the presence of superhigh magnetic fields, the values of $\Gamma$ calculated
by $\rho_{\rm e}$ derived in circular cylindrical coordinates are less than those
of $\Gamma^{'}$ calculated by $\rho_{\rm e}^{'}$ deduced in spherical coordinates,
due to the quantization of Landau levels.  Combining the relation of $\Gamma=
K(B) \Gamma^{'}$ with Landau level-superfluid modified factor $\Lambda$ yields a
modified second-order differential equation for a superhigh magnetic field $B$ and
its evolutionary timescale $t$.

2.~Calculations show that the maximum of the field's decay timescale, $t\approx
2.9507  \times 10^{6}$ yrs when $B_{\rm 0}= 3.0\times 10^{15}$ G and $T_{0}=2.60
\times 10^{8}$ K (without considering the modified Urca reactions). Assuming
different initial internal temperatures, the superhigh magnetic fields may evolve
on timescales$\sim(10^{6}\sim10^{7})$ yrs for common magnetars.

3.~On the basis of the results of the NS-SNR  association of Zhang \& Xie (2011), we
calculate the maximum initial internal magnetic fields of magnetar progenitors to be
$\sim 2.0 \times 10^{14}- 2.93 \times 10^{15}$ G, when $T_{0}\sim 2.60 \times
10^{8}$ K. If $T_{0}\sim 2.75\times 10^{8}- 1.75\times 10^{8}$~K, there still be $
B_{\rm i}\sim(10^{14}-10^{15})$~G, which are consistent with our model.

4.~By means of statistical analysis, we found that an intense and significant positive
correlation between Log$_{10}$(SNR(Age)/Spindown(Age)) and Log$_{10} B$ for magnetars,
and all AXPs associated with SNRs look older than their real ages, whereas all SGRs
associated with SNRs appear younger than they are.

5.~We tentatively investigate the equation of braking index $n$ under pure magnetodipole
radiation, and produce schematic diagrams of $\frac{dB}{dt}-~B$ and $\frac{\dot{B}}
{B^{3}}-B$ in a wide range of $B\sim 3.0\times 10^{15}\sim 10^{12}$ G. According
to our magnetar model, braking index $n$ could be correlated with both the dipole magnetic
field  and its decay rate.

6.~`The dipolar magnetic field decay plays a significant role in making a neutron
star look older' suggested by Zhang and Xie (2011) is also applicable to magnetars.
All AXPs may have experienced relatively `normal' decay of their dipole magnetic
fields, and thus have lower values of $n$ ($n < 3$) and $\dot{P}$. In contrast,
giant/intermediate flares were detected in SGRs associated with SNRs.  Giant flares
or huge glitches cause SGRs associated with SNRs spin down quickly, and make SGRs
appear younger than their real ages.

Finally, due to the very little number of magnetars associated with SNRs, the above
conclusions are tentative, and must be observationally validated.

\begin{acknowledgments}
{We thank the anonymous referee for the care in reading the manuscript and
for useful comments which help us to improve this paper substantially, Prof.
Xiang-Dong Li for his valuable suggestions on contents of this paper and helps
on language, Profs. Yong-Feng Huang and Zi-Gao Dai for their helpful comments.
This work is partly supported by Chinese National Science Foundation through grant
No.10773005, National Basic Research Program of China (973 Program 2009CB824800),
China Ministry of Science and Technology under State Key Development Program for
Basic Research (2012CB821800), Knowledge Innovation Program of CAS KJCX$_{2}$-YW
-T09, Xinjiang Natural Science Foundation No.2009211B35, the Key Directional
Project of CAS and NSFC under projects 10173020, 10673021, 10773005, 10778631 and
10903019.}
\end{acknowledgments}

\appendix
\textbf{Appendix}
\section{An important assumption on the ${}^3P_2$ neutron Cooper pairs}
The formation of Cooper pairs is a universal quantum-mechanical phenomenon of
condensation in superfluids or superconductors.  As pointed out in the original
BCS work \citep{Bardeen57}, pairing occurs basically  between fermion states in
the vicinity of the Fermi surface. Owing to pairing correlations, there is a major
change in the low-energy spectrum of the system: A finite energy gap  between its
ground state and first excited state appears, and then the system will be subjected
to a phase transition to a superfluid (or superconductor) state below a critical
temperature.

From the analysis in Papers 1-4, the ${}^3P_2$ neutron Cooper pairs can be destroyed by
the outgoing EC neutrons easily.  However, up to the present, the physics community has
not yet produced the calculation of the collision probability at which an outgoing EC
neutron destroys one ${}^3P_2$ Cooper pair, due to special circumstances inside neutron
stars, e.g., high temperatures, high-density matter, ultra-strong magnetic fields, and so on.

As a matter of fact, in the process of EC, for each degenerate species (electrons, protons,
neutrons and neutrinos), only a fraction ($\sim kT/E_{\rm F}(i)$) of particles lying in the
vicinity of the Fermi surface can effectively contribute to the EC rate, $\Gamma$. Now, let
us carry out the following evaluation: The number of neutrons participating in EC per unit
volume, $n_{\rm n}^{'}$, is computed as
\begin{equation}
 n_{\rm n}^{'}=\frac{kT}{E_{\rm F}^{'}(n)}\int \rho_{\rm n}dE_{\rm n}
 =\frac{kT}{E_{\rm F}^{'}(n)}\times \frac{8\pi\sqrt{2}m_{\rm n}^{3/2}}{h^{3}}
 \int _{E_{\rm F}^{'}(n)}^{\langle E_{\rm n}\rangle}E_{\rm n}^{\frac{1}{2}}dE_{\rm n}.
\end{equation}
As an illustration, we can arbitrarily assume $B$= 3.0 $\times 10^{15}$ G and $T$=2.78
$\times10^{8}$ K.  From Paper 4, we obtain the following relations: $\langle E_{\rm n}\rangle
=73.57$ MeV, and $E_{\rm F}^{'}(n)$=60 MeV when $B$=3.0 $\times 10^{15}$ G. Eq.(A1) gives the
value $n_{\rm n}^{'}\sim$ 2.377 $\times 10^{34}$ cm$^{-3}$ ($\rho=\rho_{\rm 0}$).  Since the
values we assumed are the possible maximum values of $B$ and $T$, the average value of $n_
{\rm n}^{'}$ will be less than this value evaluated (2.377 $\times 10^{34}$ cm$^{-3}$), obviously.
In the interior of a neutron star where the anisotropic ${}^3P_2$ neutron superfluid exists, the
neutron number density $n_{\rm n}$=1.78 $\times 10^{38}$ cm$^{-3}$ when $\rho= \rho_{\rm 0}$
\citep{Shapiro83}, which indicates that both the number of neutrons and the number of the ${}^3P_2$
neutron Cooper pairs per unit volume are far larger than the number of these newly formed (EC)
neutrons per unit volume.  Based on the above analysis, we may make a feasible assumption that
each outgoing high-energy EC neutron can destroy one ${}^3P_2$ neutron Cooper pair.
\section{ Necessary corrections and improvements in our previous work}
 For the purpose of improving our magnetar model, we have checked our previous research in an all-round way.
In this part, we make several necessary corrections in Paper 3, and provide key improvements in Paper 4.

In Paper 3, we derived the formulae for $E_{\rm F}({\rm e})$ in superhigh magnetic fields, and concluded
that the stronger the magnetic fields, the higher the electron Fermi energy becomes.  However, the coefficient
$\frac{2}{3}$ in Eq.(15) was wrongly treated as $\frac{3}{2}$ in the subsequent calculations, causing the
formulae of $E_{\rm F}({\rm e})$ deviate the actual case slightly. We honestly apologize to readers for our
mistake.  Now, the necessary corrections in Paper 3 are presented as follows: $\frac{3\pi}{B^{*}}$ in
Eqs.(16-19) must be replaced by $\frac{4\pi}{3B^{*}}$; $\frac{(3\pi)^{2}}{16B^{*}}$ in Eq.(20) in Paper 3 must be replaced
by $\frac{\pi^{2}}{4B^{*}}$; Eq.(23) in Paper 3 can be rewritten as
\begin{equation}
 E_{\rm F}({\rm e})= 43.44[\frac{Y_{\rm e}}{0.0535} \frac{\rho}{\rho_{0}}\frac{B}{B_{\rm cr}}]
^{\frac{1}{4}}{\rm MeV}   ~~~(B^{*} \geq 1)~~. (25)
\end{equation}
To our pleasure, the higher value of $Y_{\rm e}$=0.12 (see Paper 4) will be replaced by the lower value
$Y_{\rm e}$=0.0535. The later is slightly larger than the mean value of $Y_{\rm e}$ of a neutron star,
$Y_{\rm e}$=0.05, and thus is plausible.  Furthermore, the corrections of Eq.(20) and Eq.(24) in Paper 3
don't affect the calculated results of Paper 4.  Now, some values of $E_{\rm F}({\rm e})$ of Table 2 in
Paper 3 are modified, shown as in Table 4.
 \begin{table*}[t]
\small
\caption{ The relation of $E_{\rm F}({\rm e})$ and $B$.\label{tb4}}
\begin{tabular}{@{}crrrrr@{}}
\tableline
 Name   & $Y_{\rm e}$  &$E_{rm F}(1)$  & $E_{rm F}(2)$ &$E_{\rm F}(3)$& $E_{\rm F}(4)$    \\
          &          & $B^{*}$=0 &  $B^{*}$=1 &  $B^{*}$=10 &  $B^{*}$=100    \\
$ ^{56}_{26}$Fe & 0.4643  & 0.95      &\textbf{1.35}      & \textbf{3.20}      & \textbf{8.77}     \\
\tableline
$ ^{62}_{28}$Ni & 0.4516    & 2.61    & \textbf{2.48}    & \textbf{4.58}       &\textbf{12.01}    \\

$ ^{64}_{28}$Ni  & 0.4375    & 4.31   &          & \textbf{6.44}     & \textbf{13.49}    \\

$ ^{66}_{28}$Ni  & 0.4242    & 4.45    &         &\textbf{6.54}      &  NO\\

$ ^{86}_{36}$Kr & 0.4186    & 5.66     &             &        & \textbf{14.49} \\

$ ^{84}_{34}$Se  & 0.4048    & 8.49   &              &       &\textbf{18.84}  \\

$ ^{82}_{32}$Ge   & 0.3902    &11.44   &             &       & \textbf{21.75}     \\

$ ^{80}_{30}$Zn    & 0.3750    & 14.08   &            &       &\textbf{27.62}   \\

\tableline
\end{tabular}
\tablecomments{ This table is cited from Table 2 of Paper 3. All the corrections are denoted in boldface.}
\end{table*}
From the analysis in Paper 3, when $E_{F}({\rm e})\geq$~5 MeV, the second term on the right
of Eq.(20) can be ignored.  This suggest that the second term on the right of Eq.(16) also
can be ignored. Thus, the electron energy state density can be approximately expressed as
\begin{eqnarray}
N_{\rm pha}\approx &&\frac{4\pi}{3B^{*}}(\frac{m_{\rm e}c}{h})^{3}\int_{0}^{\frac{E_{\rm F}({\rm e})}
{m_{\rm e}c^{2}}}[(\frac{E_{\rm F}({\rm e})}{m_{\rm e}c^{2}})^{2}\nonumber\\
&&- 1 -(\frac{p_{z}}{m_{\rm e}c})^{2}]^{\frac{3}{2}}d(\frac{p_{z}}{m_{\rm e}c}). (26)
\end{eqnarray}
Differentiating Eq.(26) gives the following expression:
\begin{eqnarray}
\frac{dN_{\rm pha}}{dE_{\rm e}}dE_{\rm e} \simeq \frac{4\pi}{3B^{*}}(\frac{m_{\rm e}c}{h})^{3}\frac{1}{m_{\rm e}c^{2}}
[(\frac{E_{\rm F}({\rm e})}{m_{\rm e}c^{2}})^{2}-1-(\frac{E_{\rm e}}{m_{\rm e}c^{2}})^{2}]^{\frac{3}{2}}dE_{\rm e}.(27)
\end{eqnarray}
Simplifying Eq.(27) and using the relation $\rho_{\rm e}=\frac{dn_{\rm e}}{dE_{\rm e}}=\frac{dN_{\rm pha}}{dE_{\rm e}}$, we gain
\begin{eqnarray}
&&\rho_{\rm e} = \frac{4}{3}\frac{\pi}{B^{*}}(\frac{m_{\rm e}c}{h})^{3}
\frac{1}{m_{\rm e}c^{2}}[(\frac{E_{\rm F}({\rm e})}{m_{\rm e}c^{2}})^{2}-1-(\frac{E_{\rm e}}{m_{\rm e}c^{2}})^{2}]^{\frac{3}{2}}\nonumber\\
&&=\frac{1}{3B^{*}(2\pi^{2}\hbar^{3}c^{3})}\frac{1}{m_{\rm e}c^{2}}[E_{\rm F}^{2}({\rm e})- 0.261 - E_{\rm e}^{2}]^{\frac{3}{2}},(28)
\end{eqnarray}
where $m_{\rm e}c^{2}$= 0.511 MeV is used. Taking into account of gravitation redshift and utilizing Eq.(25) and Eq.(28), Eq.(16) and Eq.(19) in Paper 4 are modified as
\begin{equation}
L_{\rm X}^{\infty}= \zeta(B,T)(1-r_{\rm g}/R) \frac{dE}{dt}~~, ~~~~~~~(29)
\end{equation}
and
\begin{eqnarray}
 L_{\rm X}^{\infty} \simeq&& \Lambda(B,T)\zeta (B,T)(1 - r_{\rm g}/R)\nonumber\\
 &&\times \frac{4}{3}\pi R_{5}^{3} \frac{2\pi}{\hbar}\frac{G_{\rm F}^{2}C_{\rm V}^{2}(1+3a^{2})}{2\pi^{2}\hbar^{3}c^{3}}\frac{8\pi\sqrt{2}m_{\rm n}^{\frac{3}{2}}}{h^{3}}\nonumber\\
 &&\times \frac{(1.60 \times 10^{-6})^{9.5}}{(2\pi^{2}\hbar^{3}c^{3})3B^{*}m_{\rm e}c^{2}}
\int_{E_{\rm F}^{'}({\rm n})}^{\langle E_{\rm n}\rangle}E_{\rm n}^{\frac{1}{2}}\langle E_{\rm n}\rangle dE_{\rm n}\nonumber\\
 &&\times \int_{Q}^{E_{\rm F}({\rm e})}[E_{\rm F}^{2}({\rm e})- 0.261 - E_{\rm e}^{2}]^{\frac{3}{2}}(E_{\rm e}- Q)^{3}dE_{\rm e},(30)
\end{eqnarray}
respectively, where $L_{\rm X}^{\infty}$ is the apparent soft X-ray luminosity measured by a
distant observer or the redshifted soft X-ray luminosity; $r_{\rm g}= 2GM/c^{2}= 2.95M/M_{\rm {\odot}}$ km
is the Schwarzschild radius (we assume $R =10^{6}$ cm and $M= 1.4M_{\rm {\odot}}$ for a canonical magnetar).
We introduce the parameter $\phi(B, T)$ to denote $\Lambda(B, T)\zeta (B,T)$ in Eq.(30), the value of $\phi(B,T)$
of a magnetar can be evaluated by combining Eq.(B6) with Table 4 of Paper 4.
\begin{figure}[th]
\centering
  \vspace{0.5cm}
  \includegraphics[width=8.1cm,angle=360]{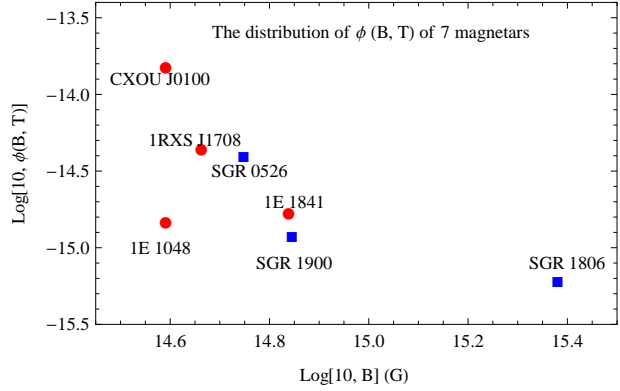}
  \caption{The distribution of $\phi(B,T)$ of 7 canonic magnetars when $B_{0}= 3.0 \times 10^{15}$ G and $T_{0}= 2.6 \times 10^{8}$ K. The range of $B$ is assumed to be ($1.80 \times 10^{14}\sim 3.0 \times 10^{15}$ G, arbitrarily. $Circles$ and
$squares$ are for AXPs and SGRS, respectively.}
  \label{fg:Fig.9}
  \end{figure}
Fig.9 shows the distribution of $\phi(B,T)$ of 7 canonic magnetars, whose persistent soft X-ray luminosities
should not be less than their rotational energy loss rates $dE/dt$ (Paper 4). It is worth noting
that the fitting curve of $\phi(B, T)$ as a function of $B$ and $T$ cannot be obtained
from Fig.9  because of the very little number of canonic magnetars.  For each canonical
magnetar with a given soft X-ray luminosity, its value of $\phi$ is mainly determined by
$B$, and is insensitive to $T$, though $\phi$ is a function of $B$ and $T$. The mean
value of $\phi(B,T)$ of 7 magnetars is calculated to be $2.54\times 10^{-15}$ by using
the expression of $\langle \phi\rangle= \frac{\sum B_{i}\phi_{i}}{\sum B_{i}}$. Combining
Eq.(7) with $\langle\phi\rangle$ gives
\begin{eqnarray}
\phi(B, T)\approx &&\frac{2.54\times 10^{-15}}{3.198\times 10^{-14}}(2.60388\times 10^{-14}\nonumber\\
&&+~1.98233\times 10^{-30}B)(\frac{T_{0}}{2.6\times10^{8}\rm K})^{4}\nonumber\\
&&\times exp[\frac{-0.048{\rm MeV}}{k}(\frac{1}{T_{0}}-\frac{1}{2.6\times10^{8}{\rm K}})]\nonumber\\
&&=(2.068\times 10^{-15}+~1.5747\times 10^{-31}B)\nonumber\\
&&\times (\frac{T_{0}}{2.6\times10^{8}\rm K})^{4}\nonumber\\
&&\times exp[\frac{-0.048{\rm MeV}}{k}(\frac{1}{T_{0}}-\frac{1}{2.6\times10^{8}{\rm K}})].(31)
\end{eqnarray}
Inserting Eq.(31) into Eq.(30), we calculate the values
of $L_{\rm X}^{\infty}$ in superhigh magnetic fields, $L_{\rm X}^{\infty}\sim (5.457 \times 10^{28}-3.834\times
10^{37})$ erg~s$^{-1}$, corresponding to $B\sim (1.8 \times 10^{14}- 3.0 \times 10^{15})$ G.  Furthermore, the
calculated results of $L_{\rm X}^{\infty}$ are compared with those of $L_{\rm X}$ of Paper 4, shown as in Fig.10
\begin{figure}[th]
\centering
  \vspace{0.5cm}
  \includegraphics[width=8.1cm,angle=360]{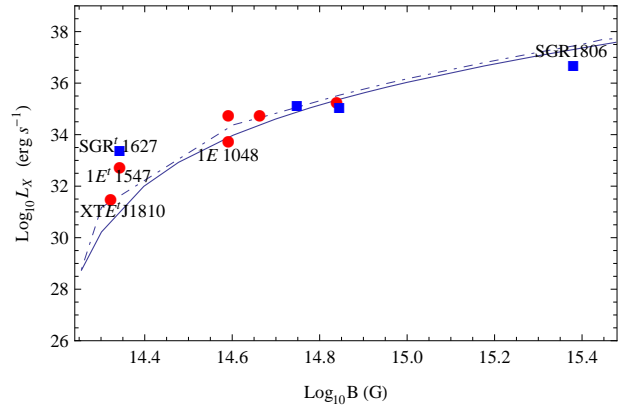}
  \caption{The diagrams of soft X-ray luminosity
as a function of magnetic field strength $B$ when
$B_{0}=3.0\times 10^{15}$ G and $T_{0}= 2.6\times 10^{8}$ K.
The range of $B$ is assumed to be ($1.80 \times 10^{14}\sim 3.0
\times 10^{15}$ G, arbitrarily. $Circles$ and $squares$ are for AXPs
and SGRS, respectively. $Solid line$ and $dot-dashed line$ are for the
modified luminosities and unmodified luminosities (Paper 4), respectively.}
  \label{fg:Fig.10}
\end{figure}
From Fig.10, the values of $L_{\rm X}^{\infty}$ are slightly less than those of $L_{\rm X}$, the main reason
for this is that the factor of gravitation redshift is considered in calculating $L_{\rm X}^{\infty}$.


\begin{thebibliography}{}
\bibitem[Aguilera et~al.(2008)]{Aguilera08} Aguilera, D. N., Pons, J. A., Miralles, J. A., 2008,~\apjl, 673, L167
\bibitem[Aharonian et~al.(2008)]{Aharonian08}Aharonian, F., Akhperjanian, A. G., Barres, de Almeida, U; et~al., 2008,~\aap,486,829
\bibitem[Allen \& Horvath(1997)]{Allen97} Allen, M. P., Horvath, J. E., 1997, \apj, 488, 409
\bibitem[Allen \& Horvath(2004)]{Allen04} Allen, M. P., Horvath, J. E., 2004, \apj, 616, 346
\bibitem[Alpar et~al.(2001)]{Alpar01}Alpar, M. A., Ankay, A., Yazgan, E., 2001, \apjl, 577, L61
\bibitem[Arras et~al.(2004)]{Arras04}Arras, P., Cumming, A., Thompson, C., 2004,~\apjl, 608, L49
\bibitem[Bardeen et~al.(1957)]{Bardeen57} Bardeen, J., Cooper, L. N., Schrieffer, J.R.,1957, Phys. Rev. 108, 1175
\bibitem[Blandford \& Romani(1988)]{Blandford88}Blandford, R. D., Romani, R. W., 1988, \mnras, 234, 57
\bibitem[Camilo et~al.(2007)]{Camilo07}Camilo, F., Ransom, S. M., Halpern, J. P., Reynolds, J., 2007, \apjl, 666, L93
\bibitem[Chen \& Li(2006)]{Chen06} Chen, W. C., Li, X. D., 2006, \aap, 450, L1
\bibitem[Corbel et~al.(1999)]{Corbel99} Corbel, S., Chapuis, C., Damc, T. M., Durouchoux, P.,1999, \apjl, 526, L29
\bibitem[Duncan \& Thompson (1992)]{Duncan92} Duncan R. C., Thompson C., 1992, \apj, 392, L9
\bibitem[Duncan \& Thompson(1996)]{Duncan96}Duncan R. C., Thompson C., In: Rothschild
  R.E., Lingenfelter R.E. (eds.) High-Velocity Neutron Stars and Gamma-Ray Bursts. AIP
  Conference Proc., Vol. 366, p. 111. AIP Press, New York (1996)
\bibitem[Duncan(2000)]{Duncan00}Duncan, R. C., in Fifth Hunstville Gamma-Ray Burst Symposium, AIP
Conference Proceedings No. 526, ed. R. M. Kippen, R. S. Mallozzi, \& G. J.
Fishman, American Institute of Physics, New York (2000), arXiv:astro-ph/0002442
\bibitem[Elgar${\o}$y et~al.(1996)]{Elgar96} Elgar${\o}$y, ${\O}$., et~al., 1996, \prl, 77, 1482
\bibitem[Fahlman \& Gregory(1981)]{Fahlman81} Fahlman, G. G., Gregory, P. C., 1981, \nat, 293, 202
\bibitem[Gaensler et~al.(1999)]{Gaensler99}Gaensler, B. M., Gotthelf, E. V., Vasisht, G., 1999,\apjl, 526, L37
\bibitem[Gaensler et~al.(2001)]{Gaensler01}Gaensler, B. M., Gotthelf, E. V., Vasisht, G., 2001,\apj, 559, 963
\bibitem[Gelfand \& Gaensler(2007)]{Gelfand07}Gelfand J. D., Gaensler, B. M., 2007, \apj, 667, 1111,arXiv:0706.1054
\bibitem[Green (1989)]{Green89} Green, D. A., 1989, \mnras, 238, 737
\bibitem[Gao et~al.(2011a)]{Gao11a} Gao, Z. F., Wang, N., Yuan, J. P., et al., 2011a, \apss,332, 129
\bibitem[Gao et~al.(2011b)]{Gao11b} Gao, Z. F., Wang, N., Yuan, J. P., et al., 2011b, \apss, 333, 427
\bibitem[Gao et~al.(2011c)]{Gao11c} Gao, Z. F., Wang, N., Song, D. L., et al., 2011c, \apss, 334, 281
\bibitem[Gao et~al.(2011d)]{Gao11d} Gao, Z. F., Peng, Q. H., Wang, N., et al., 2011d, \apss, 336, 427
\bibitem[Goldreich \& Reisenegger (1992)]{Goldreich92}Goldreich, P., Reisenegger, A., 1992, \apj, 395, 250
\bibitem[Gunn \& Ostriker(1970)]{Gunn70}Gunn, J. E., Ostriker, J. P., 1970, \apj, 160, 979
\bibitem[Halpern \& Gotthelf(2010)]{Halpern10}  Halpern, J. P.,  Gotthelf, E. V.,  2010, \apj, 725, 1384
\bibitem[Han(1997)]{Han97}Han, J. L., 1997, \aap, 489, 485
\bibitem[Heyl \& Hernquist(1997)]{Heyl97}Heyl, J. S., Hernquist, L., 1997, \apjl, 489, L67
\bibitem[Heyl \& Kulkarni(1998)]{Heyl98}Heyl, J. S., Kulkarni, S. R., 1998, \apjl, 506, L61
\bibitem[Hurley et~al.(1999)] {Hurley99} Hurley, K., Kouveliotou, C., Woods, P., et~al., 1999, \apjl, 510, L107
\bibitem[Horiuchi et~al.(2008)] {Horiuchi08}Horiuchi, S., Suwa, Y., Takami, H.,  et~al., 2008, \mnras, 391, 1893
\bibitem[Horvath \& Allen (2011)]{Horvath11}Horvath, J. E., Allen, M. P., 2011, Res. Astron. Astrophys. 11, 625 
\bibitem[Ibrahim et~al.(2001)]{Ibrahim01} Ibrahim, A. I., Strohmayer, T. E., Woods, P. M., et~al., 2001, \apj, 558, 237
\bibitem[Ibrahim et~al.(2004)]{Ibrahim04} Ibrahim, A. I., Markwardt, C. B.,  Swank, J. H, et~al., 2004, \apjl, 609, L1
\bibitem[Klose  et~al.(2004)]{Klose04}Klose, S., Henden, A. A., Geppert, U., et~al. 2004, \apjl, 609, L13, arXiv:astro-ph/0405299
\bibitem[Kouveliotou  et~al.(1994)]{Kouveliotou94} Kouveliotou, C., Fishman, G. J., Meegan, C. A., et~al., 1994, \nat, 368, 125
\bibitem[Kulkarni \& Frail (1993)]{Kulkarni93}Kulkarni, S. R., Frail, D. A.,1993,\nat, 365, 33
\bibitem[Kulkarni et~al.(2003)]{Kulkarni03}Kulkarni, S. R., Kaplan, D. L., Marshall, H. L., et~al., 2003, \apj, 585, 948
\bibitem[Lenters et~al.(2003)]{Lenters03}Lenters, G. T., Woods, P. M., Goupell, J. E., et~al., 2003, \apj, 587, 761
\bibitem[Lin \& Zhang (2004)]{Lin04} Lin, J. R., Zhang, S. N., 2004, \apjl, 615, L133
\bibitem[Manchester \& Taylor (1977)]{Manchester77}Manchester, R. N., Taylor, J. H.,: Pulsars. San Francisco, CA(USA): W. H. Freeman, 281p
\bibitem[Marsden et~al.(1999)]{Marsden99} Marsden, D., Rothschild, R. E., Lingenfelter, R. E., 1999, \apjl, 520, l107
\bibitem[Marsden et~al.(2001)]{Marsden01} Marsden, D.,  Lingenfelter, R. E.,  Rothschild, R. E., et~al., 2001, \apj, 550, 397
\bibitem[Mazets et~al.(1999)]{Mazets99} Mazets, E. P., Aptekar, R. L., Butterworth, P. S., 1999, \apjl, 519, L151
\bibitem[Mereghetti et~al.(2000)]{Mereghetti00} Mereghetti, S., Cremonesi, D., Feroci, M., Tavani, M., 2000, \aap, 361, 240
\bibitem[Menou et~al.(2001)]{Menou01}Menou, K., Perna, R., Hernquist, L., 2001, \apjl, 554, L63
\bibitem[Mereghetti et~al.(2005)]{Mereghetti05} Mereghetti, S., G$\ddot{o}$tz, D., Mirabel, I. F., et~al., 2005, \aap, 433, L9
\bibitem[Mereghetti(2008)]{Mereghetti08} Mereghetti, S., 2008, \aapr, 15, 225
\bibitem[Narayan \& Ostriker(1990)]{Narayan90}Narayan, R., Ostriker, J. P., 1990, \apj, 352, 222
\bibitem[Ostriker \& Gunn(1969)]{Ostriker69}Ostriker, J. P., Gunn, J. E., 1969, \apj, 157, 1395
\bibitem[Peng et~al.(1982)]{Peng82} Peng, Q. H., Huang, K. L., Huang, J. H., 1982, \aap, 107, 258
\bibitem[Peng \& Tong(2007)]{Peng07} Peng, Q. H., Tong  H., 2007, \mnras, 378, 159
\bibitem[Peng \& Tong (2009)]{Peng09} Peng Qiu He., Tong Hao., arXiv: 0911.2066v1 [astro-ph.HE]
11 Nov 2009, $10^{th}$ Symposium on Nuclei in the Cosmos, 27 July-1 August 2008 Mackinac Island, Michigan,USA
\bibitem[Pons et~al.(2007)]{Pons07}Pons, J. A., Link, B., Miralles, J. A., \& Geppert, U., 2007, Phys. Rev. Lett, 98,
071101
\bibitem[Pons et~al.(2009)]{Pons09}Pons, J. A., Miralles, J. A., Geppert, U., 2009, \aap, 496, 207
\bibitem[Pons \& Rea(2012)]{Pons12}Pons, J. A., Rea, N., 2012,\apjl, 750, 6
\bibitem[Rheinhardt \& Geppert(2003)]{Rheinhardt03}Rheinhardt, M., Geppert, U., 2003, \prl, 88, L101
\bibitem[Ridley \& Lorimer(2010)]{Ridley10} Ridley, J. P., Lorimer, D. R., 2010, \mnras, 404, 1081
\bibitem[Rho \& Petre (1997)]{Rho97} Rho, J., Petre, R., 1997, \apj,484, 828
\bibitem[Sanbonmatsu \& Herfand(1992)]{Sanbonmatsu92}Sanbonmatsu, K. Y., Herfand, D. J., 1992,\apj, 104, 2189
\bibitem[Shapiro \& Teukolsky (1983)]{Shapiro83}Shapiro, S. L., Teukolsky, S. A., 1983, `Black holes,white
        drarfs,and neutron stars' John Wiley \& Sons, New York
\bibitem[Shull et~al.(1989)]{Shull89} Shull, J. M., Fesen, R. A., Saken, J. M., 1989, \apj, 346, 860
\bibitem[Ternov et~al.(1965)]{Ternov65}Ternov, I., Lysov, B., Korovina, L., 1965, Moscow Univ. Phys. Bull. 5, 58
\bibitem[Thompson \& Duncan (1993)]{Thompson93} Thompson, C., Duncan, R. C., 1993, \apj, 543, 340
\bibitem[Thompson \& Duncan(1996)]{Thompson96} Thompson, C., Duncan, R. C., 1996, \apj, 473, 322
\bibitem[Thompson et~al.(2000)]{Thompson00}Thompson, C., Duncan, R. C., Woods, P. M., 2000, \apj, 543, 340
\bibitem[Thompson et~al.(2002)]{Thompson02}Thompson, C., Lyutikov, M.,  Kulkami, S. R., 2002, \apj, 574, 332
\bibitem[van Paradijs et~al.(1995)]{van95}  van Paradijs, J., Taam, R. E., van den Heuvel, E. P. J., 1995, \aap, 299, 41
\bibitem[ Vasisht et~al.(1994)]{Vasisht94} Vasisht, G., Kulkarni, S. R., Frail, D. A., et~al., 2000, \apjl, 431, L35
\bibitem[ Vasisht \& Gotthelf (1997)]{Vasisht97} Vasisht, G.,  Gotthelf, E. V., 1997, \apjl, 486, L129
\bibitem[ Vasisht et~al.(2000)]{Vasisht00} Vasisht, G., Gotthelf, E. V., Torri, K., et~al., 2000, \apjl, 542, L49
\bibitem[Wachter et~al.(2004)]{Wachter04}Wacheter, S., Patel, S., Kouveliotou, C., et~al., 2004, \apj, 615, 887
\bibitem[White \& Marshall(1984)]{White84}White, N. E., Marshall, F. E., 1984,\apj, 281, 354
\bibitem[Wu et~al (2003)]{Wu03} Wu, F., Xu, R. X., Gil, J., 2003, \aap, 409, 641
\bibitem[Xu \& Qiao (2001)]{Xu01} Xu, R. X., Qiao, G. J., 2001, \apjl, 561, L85
\bibitem[Xu et~al.(2006)]{Xu06} Xu, R. X., Tao, D. J., Yang, Y., 2006, \mnras, 373, L85
\bibitem[Yakovlev et al. (2001)]{Yakovlev01} Yakovlev, D. G., Kaminker A. D., Gnedin O. Y., et al., 2001, \physrep, 354, 1
\bibitem[Zhang et~al.(2000)]{Zhang00} Zhang, B., Xu, R. X.,  Qiao, G. J., 2000, \apjl, 545, L127
\bibitem[Zhang \& Xie(2011)]{Zhang11} Zhang, S. N., Xie, Y.,  9th Pacific Rim Conference on Stellar Astrophysics, Lijiang, China in 14-20 April 2011.,
ASP Conference Series, Vol. 451. Edited by S. Qain, K. Leung, L. Zhu, and S. Kwok. San Francisco: Astronomical Society of the Pacific, p.231 (2011).
arXiv:1110.3154v1[astro-ph.HE]


\end{thebibliography}
\end{document}